\documentclass[a4paper, amsfonts, amssymb, amsmath, reprint, showkeys, footinbib, ,twoside,superscriptaddress,floatfix,longbibliography]{revtex4-1}

\usepackage{graphicx} %
\usepackage{amsmath}
\usepackage{comment}
\usepackage{booktabs}
\usepackage{hyperref}
\usepackage{url}
\usepackage{adjustbox}
\usepackage{array}
\usepackage{xcolor}
\usepackage[USenglish]{babel}

\newcommand{\figLabelCapt}[1]{\textbf{\MakeLowercase{{#1}}}}
\newcommand{\refSub}[2]{\hyperref[#2]{\ref{#2}\figLabelCapt{#1}}}

\newcommand{\br}[1]{\mathbf{r}}
\newcommand{\bk}[1]{\mathbf{k}}

\begin{document}

\title{Learning charges and long-range interactions from energies and forces}

\author{Dongjin Kim}
\thanks{These authors contributed equally.}
\affiliation{Department of Chemistry, UC Berkeley, California 94720, United States}

\author{Daniel S. King}
\thanks{These authors contributed equally.}
\affiliation{Bakar Institute of Digital Materials for the Planet, UC Berkeley, California 94720, United States}

\author{Peichen Zhong}
\thanks{These authors contributed equally.}
\affiliation{Bakar Institute of Digital Materials for the Planet, UC Berkeley, California 94720, United States}

\author{Bingqing Cheng}
\email{bingqingcheng@berkeley.edu}
\affiliation{Department of Chemistry, UC Berkeley, California 94720, United States}
\affiliation{Bakar Institute of Digital Materials for the Planet, UC Berkeley, California 94720, United States}
\affiliation{The Institute of Science and Technology Austria, Am Campus 1, 3400 Klosterneuburg, Austria}

\date{\today}

\begin{abstract}
Accurate modeling of long-range forces is critical in atomistic simulations, as they play a central role in determining the properties of materials and chemical systems. However, standard machine learning interatomic potentials (MLIPs) often rely on short-range approximations, limiting their applicability to systems with significant electrostatics and dispersion forces. 
We recently introduced the Latent Ewald Summation (LES) method, which captures long-range electrostatics without explicitly learning atomic charges or charge equilibration. 
Extending LES, we incorporate the ability to learn physical partial charges, encode charge states, and the option to impose charge neutrality constraints. 
We benchmark LES on diverse and challenging systems, including charged molecules, ionic liquid, electrolyte solution, polar dipeptides, surface adsorption, electrolyte/solid interfaces, and solid-solid interfaces.
Our results show that LES can effectively infer physical partial charges, dipole and quadrupole moments, as well as achieve better accuracy compared to methods that explicitly learn charges. 
LES thus provides an efficient, interpretable, and generalizable MLIP framework for simulating complex systems with intricate charge transfer and long-range forces. 
\end{abstract}

\maketitle

\section{Introduction}

The accurate incorporation of long-range interactions in atomistic simulations of materials and chemical systems remains a fundamental challenge~\cite{french2010long}. 
Early approaches to address this issue included the cluster expansion formalism for crystalline lattices~\cite{efficient_CE}, parameterization of classical force fields with fixed charges~\cite{kirby2019charge}, and charge equilibration schemes~\cite{siepmann1995influence}, among others.

The proliferation of machine learning interatomic potentials (MLIPs)~\cite{keith2021combining,unke2021machine}, which learn surrogate potential energy surfaces from quantum mechanical reference calculations of atomic configurations, has further emphasized the need for accurately accounting for long-range interactions. 
Most established MLIP frameworks rely on short-range approximations, assuming that the energy contribution of each atom is determined by its local atomic environment. While this assumption enables computationally efficient linear scaling with respect to system size, it poses significant limitations for systems where long-range interactions, such as electrostatics, play a critical role.
These limitations are particularly evident in systems involving electrochemical interfaces~\cite{niblett2021learning}, charged molecular dimers~\cite{grisafi2019incorporating,huguenin2023physics}, ionic~\cite{zhang2022deep} and polar materials~\cite{monacelli2024electrostatic}, and scenarios involving varying charge states or long-range charge transfer~\cite{ko2021fourth}. 

One option is to predict effective partial charges of each atom, which are then used to determine long-range electrostatics~\cite{unke2019physnet,ko2021fourth,gao2022self,sifain2018discovering,gong2024bamboo,shaidu2024incorporating}.
For example, the third-generation HDNNP (3G-HDNNP)~\cite{ko2021fourth} contains electrostatic
interactions based on local environment-dependent charges represented by atomic neural networks.
To improve upon that,
the fourth-generation high-dimensional neural network potential (4G-HDNNPs)~\cite{ko2021fourth} predicts the electronegativities of each atom and then uses a charge equilibration scheme~\cite{rappe1991charge} to assign the charges. 
3G-HDNNPs and 4G-HDNNPs are trained directly to reproduce atomic partial charges from reference quantum mechanical calculations, although partial charges are not physically observable and their values depend on the specific partitioning scheme used~\cite{sifain2018discovering}.
Another approach is to learn the maximally localized Wannier centers (MLWCs) for insulating systems:
the deep potential long-range (DPLR) model~\cite{zhang2022deep} computes the long-range electrostatics using spherical Gaussian charges associated with the nuclei and the
average positions of the MLWCs predicted via a
Deep Wannier (DW) deep neural network model based on the local chemical environment~\cite{zhang2022deep}.
The charges of these MLWCs are based on the number of valence electrons of each element.
A similar method is the self-consistent field neural network (SCFNN)~\cite{gao2022self}, which predicts the electronic response via the position of the MLWCs.

There are a few other methods that do not explicitly learn the atomic charges~\cite{yu2022capturing,kosmala2023ewald,grisafi2019incorporating,huguenin2023physics,faller2024density,monacelli2024electrostatic}.
For example, the Ewald message-passing method~\cite{kosmala2023ewald} employs a learnable frequency filter in the reciprocal space to generate a long-range message for each atom during the message-passing step.
LODE~\cite{grisafi2019incorporating,huguenin2023physics,loche2024fast} computes the potential field generated by all the atoms in the system in the reciprocal space via Ewald summation, and then featurizes such field near a central atom up to some cutoff radius to form the long-range descriptors. 
The density-based long-range descriptor~\cite{faller2024density} follows a similar procedure, but the global atomic density itself is used instead of the field.

Recently, we introduced the Latent Ewald Summation (LES) method~\cite{cheng2024latent}. 
LES decomposes the total potential energy into short-range and long-range components. 
Hidden variables--interpreted as “latent charges”--are predicted from local atomic features without reference to specific charge definitions. These latent charges are then used to predict the long-range potential via an Ewald summation.
LES can be combined with any short-ranged MLIP architectures (e.g. HDNNP~\cite{behler2007generalized}, Gaussian Approximation Potentials (GAP)~\cite{bartok2010gaussian}, Moment Tensor Potentials (MTPs)~\cite{shapeev2016moment}, atomic cluster expansion (ACE)~\cite{drautz2019atomic}) and MPNN (e.g.,
NequIP~\cite{batzner20223}, MACE~\cite{batatia2022mace}).
We combine LES with Cartesian atomic cluster expansion (CACE)~\cite{cheng2024cartesian}, and refer to the standard short-ranged CACE as CACE-SR, and the combined long-range potential as CACE-LR.

In this paper, we provide a comprehensive exploration of the LES framework, detailing its theoretical foundation, new possible extensions, and application to a range of test systems. 
Importantly, we show that, when limited to a single charge channel, the LES framework is able to infer physical partial charges and dipole moments by just learning from reference energy and forces.
In Ref.~\cite{cheng2024latent}, LES was compared to other LR methods such as LODE~\cite{grisafi2019incorporating,huguenin2023physics} and density-based long-range descriptor~\cite{faller2024density} that do not explicitly learn charges.
Here, we further compare LES to existing methods that incorporate long-range interactions via explicit charge learning and show that LES achieves superior performance.

\section{Theory}

We first briefly recap LES~\cite{cheng2024latent}, and then make an explicit connection between LES and physical charges. Finally, we briefly demonstrate how different global charge states can be encoded in the LES framework.

\paragraph{Range separation}
The total potential energy of a system with $N$ atoms is split into short-range (SR) and long-range (LR) components,
$E = \sum_{i=1}^{N} E^{sr}(B_i) + E^{lr}$.
The short-range energy is the sum of atomic energies, each depending on local $B$ features of atom $i$.
The $B$ features can be local atomic environment descriptors such as ACE~\cite{drautz2019atomic},
or learned features in message passing neural networks (MPNNs)~\cite{schutt2017schnet,batzner20223,deng2023chgnet,haghighatlari2022newtonnet}.
For the long-range part, a multilayer perceptron with parameters \( \phi \) maps the invariant features of each atom \( i \) to a hidden variable:
\begin{equation}
    q_i = Q_{\phi}(B_i).
\end{equation}
In general, $q$ can be multi-dimensional to represent the generalized long-range interactions. When $q$ is restricted to be one-dimensional, it can be interpreted as the atomic charge as we discuss later.

Suppose that the potential-generating field by a single particle with unity latent variable is proportional to $ u(\mathbf{r}) = | \mathbf{r} |^{-p}$,
with $p$ being a fixed exponent.
Following the standard range-separation formalism~\cite{williams1971accelerated}, one can express short-range and long-range interactions by multiplying the interaction by a convergence function $\varphi(r)$ with $\varphi(0)=1$ decreasing rapidly to zero as $r$ increases:
\begin{multline}
    E_p = E_p^{sr} - E_p^{self}  + E_p^{lr}= \\
    \frac{1}{2} \sum_{i \neq j} q_i q_j r_{ij}^{-p} \varphi(r_{ij}) 
     - \frac{1}{2} \sum_{i=1}^N q_i^2 \lim_{r\rightarrow 0}r^{-p} [1 - \varphi(r)]
     \\
   + \frac{1}{2} \sum_{i=1}^{N} \sum_{j=1}^N q_i q_j r_{ij}^{-p} [1 - \varphi(r_{ij})].
    \label{eq:e_tot}
\end{multline}
Both $E_p^{sr}$ and $E_p^{self}$ are short-ranged in nature and can be described by the short-ranged MLIP based on the local features.

\paragraph{Long-range energy}
For $p=1$, which corresponds to electrostatics,
one choice for the convergence function can be expressed as the complimentary error function $\varphi(r) = \mathrm{erfc}(\frac{r}{\sqrt{2}\sigma})$.
For isolated systems without periodic boundary conditions, one can compute the $E_p^{lr}$ term directly in the real space based on enumerating pairwise distances between atoms.
For periodic systems,
the corresponding long-range electrostatics can be computed in the reciprocal space as
\begin{equation}
        E_1^{lr} 
    = \dfrac{2\pi}{V} \sum_{0<k<k_c}
    \dfrac{1}{k^2}
    e^{-\sigma^2 k^2/2} |S(\mathbf{k})|^2,
    \label{eq:e_lr}
\end{equation}
where the structure factor $S(\mathbf{k})$ of the hidden variable is defined as
\begin{equation}
    S(\mathbf{k}) = \sum_{i=1}^N q_i e^{i\mathbf{k}\cdot\mathbf{r}_i}.
    \label{eq:sfactor}
\end{equation}
The omission of the $k=0$ term in eq.~\eqref{eq:e_lr} means the tinfoil boundary condition is applied.
The detailed derivations and the case for $p=6$ which corresponds to London dispersions can be found in the Appendix~\ref{appdix:lr_decay}.

\paragraph{Learning charges from energy and forces}
When training the MLIP, the total potential energy $E$, interatomic forces $\mathbf{F}_i = -\partial E / \partial \mathbf{r}_i$, and sometimes virial stress are fitted to the reference values from the dataset. 
In LES, unlike methods that explicitly learn partial charges, the hidden variables $q$ are hypothesized to represent flexible atomic charges when the physical electrostatic constant $1/4\pi \epsilon_0$ is included. In particular, when LES is limited to a single charge channel, we find that the charge used to compute the long-range energy in Eq.~\eqref{eq:e_lr} is physically meaningful and can be used to predict physical observables such as the dipole moment. However, it is noted that because the structure factor is squared in Eq.~\eqref{eq:e_lr}, the predicted charges do not distinguish the charge parity, as the total energy stays the same if all signs of the charges are flipped. In practice, it is easy to
``unflip'' the signs of atomic charges
based on the known electronegativity of elements.

We note the success of the LES method in predicting charge locally while computing energy globally. This choice reflects the generally nearsightedness of electron matter~\cite{prodan2005nearsightedness}.
Additionally, as LES learns the charges via the energy and forces, its learning is flexible to arbitrary charge distributions (e.g., different oxidation states) as long as they have an impact on the energy in the training set. Indeed, the LES approach proves appropriate for a wide range of systems such as electrolyte/electrodes, charged molecules, and doped surfaces, as we will show in the examples. 
However, it is important to note that while the local charge assumption works well empirically, it lacks theoretical guarantees and may encounter limitations in specific edge cases, such as systems involving long-range charge transfer.

\paragraph{Charge neutrality condition}
Empirically, in the examples below and in previous work ~\cite{cheng2024latent} we have found it unnecessary to explicitly enforce charge neutrality or fixed total charge state in the training process of LES. In practice, we have found that the sum of $q$ is usually close to the total charges for both neutral and charged systems without enforcing neutrality.
Additionally, any residual difference is treated as a uniform background charge, which does not affect the total energy as the $k=0$ term is omitted in the reciprocal space computation of electrostatic interactions in Eq.~\eqref{eq:e_lr}.
In all the examples we have tested, we did not observe any loss of accuracy or artifacts due to the lack of charge equilibration. In contrast, for the ML models that explicitly learn charges such as 3G-HDNNP~\cite{ko2021fourth}, the lack of charge equilibration may result in dramatically larger errors, and sometimes pathological behaviors were observed for systems involving charge transfer and change of charge states.

Nevertheless, we note that it is possible to fix the total charge while avoiding the charge equilibration. One possibility is to add the following penalty term to the total potential energy $E$:
\begin{equation}
    E^{\lambda} = \lambda \left(Q - \sum_{i=1}^N q_i\right)^2,
\end{equation}
where the positive constant $\lambda$ can be understood as a Lagrangian multiplier,
and $Q$ is the referenced total charge of the system.
Although we do not use this scheme in any of the examples, we provide it here for future use case.

\paragraph{Different charge states}
In a standard MLIP, the atomic features $B_i$ depend on the chemical elements and the coordinates of the atoms surrounding atom $i$, and are agnostic to the charge or oxidation state.
This means that two systems with identical atomic positions but different net charges $Q$ will have degenerate features.
Although this degeneracy does not affect the training or prediction for systems with a fixed net charge, it can cause problems when handling systems with varying charge states simultaneously.
To resolve this, in training sets containing multiple net charges (only one of the examples below, Ag$_3^+$/Ag$_3^-$) we concatenate the total charge $Q$ of the system with the local atomic features $B_i$,
$B_i \oplus Q$, and use this combined feature as the input for predicting short-range atomic energies and local hidden variables.

\section{Examples}

\subsection{Random charges}

\begin{figure}
  \centering
\includegraphics[width=0.99\linewidth]{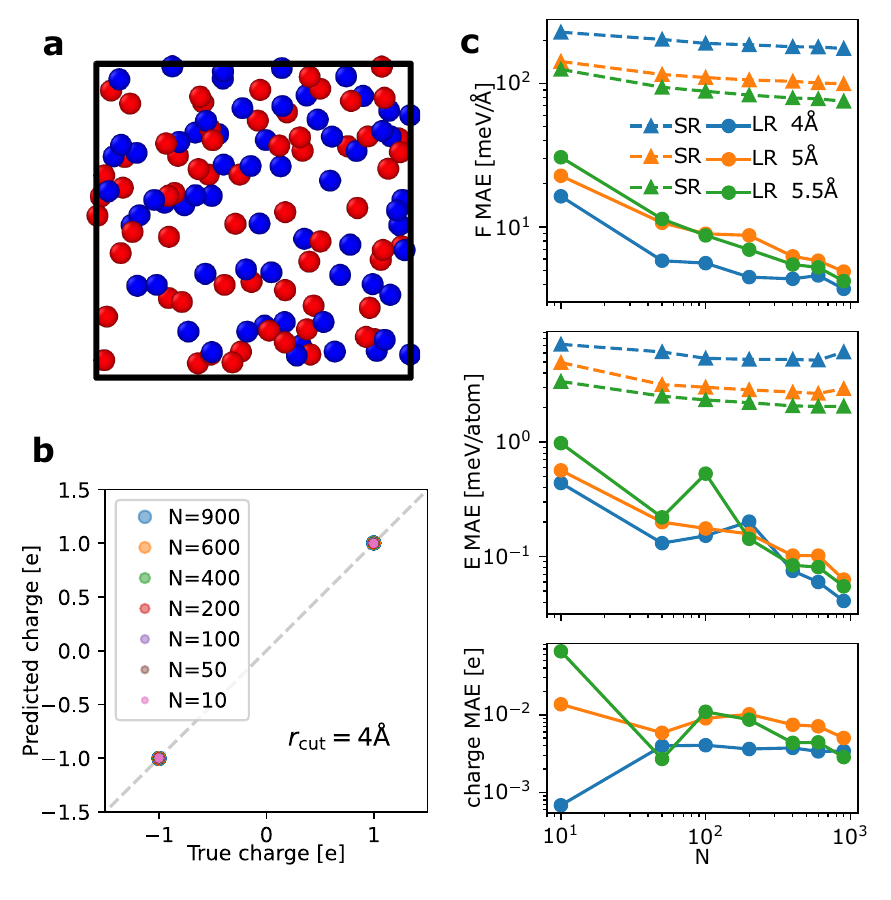}
    \caption{
    \figLabelCapt{a} A configuration of gas made of point charges.
    \figLabelCapt{b} Comparison of the true and the predicted charges for the CACE-LR models with a cutoff radius of $4$~\AA{} and trained on $N$ configurations.
    \figLabelCapt{c} The mean absolute errors (MAEs) on energy, forces, and charges for short-range (SR) and long-range (LR) models trained using different $N$ numbers of samples.
    }
    \label{fig:randomcharge}
\end{figure}

As an initial test, a gas of point charges was constructed. 
As shown in Fig.~\ref{fig:randomcharge}a, each configuration consists of 128 atoms, with 64 carrying a positive charge of $+1$e and the remaining 64 carrying a negative charge of $-1$e. 
The atoms interact through the Coulomb potential and the repulsive component of a Lennard-Jones potential. This benchmark aims to evaluate the learning efficiency of the LES framework and assess whether the correct atomic charges can be accurately learned. 
Unlike in density functional theory (DFT), where the precise values of partial charges depend on the chosen definition, the charges in this system are unambiguously defined.

For the short-range component, we employed CACE with different cutoff distances of $r_\mathrm{cut} = 4$~\AA, $5$~\AA, and $5.5$~\AA. For the long-range interactions, we used a one-dimensional $q$ with $\sigma = 1$~\AA{} in the Ewald summation, without enforcing a net charge constraint. Fig.~\ref{fig:randomcharge}b presents the parity plot of the CACE-LR model with $r_\mathrm{cut} = 4$~\AA{}, comparing the true and predicted charges (after ``unflipping'' the charge parity) for various numbers of training samples. Remarkably, even with just 10 training configurations, the predicted charges are nearly exact.

Fig.~\ref{fig:randomcharge}c illustrates the learning curves for the mean absolute errors (MAEs) in energy, forces, and charges, using short-range (SR) and long-range (LR) models with different cutoffs. The SR models exhibit slow learning and significant errors for this dataset, with performance improving as $r_\mathrm{cut}$ increases. In contrast, the LR models achieve errors more than an order of magnitude lower, with learning efficiency improving as $r_\mathrm{cut}$ decreases.
This example highlights that, unlike the typical behavior of SR machine learning interatomic potentials (MLIPs), long-range potentials achieve more efficient learning with appropriately small $r_\mathrm{cut}$ values.

\subsection{Electrolyte solutions}
\begin{figure}
  \centering
\includegraphics[width=0.9\linewidth]{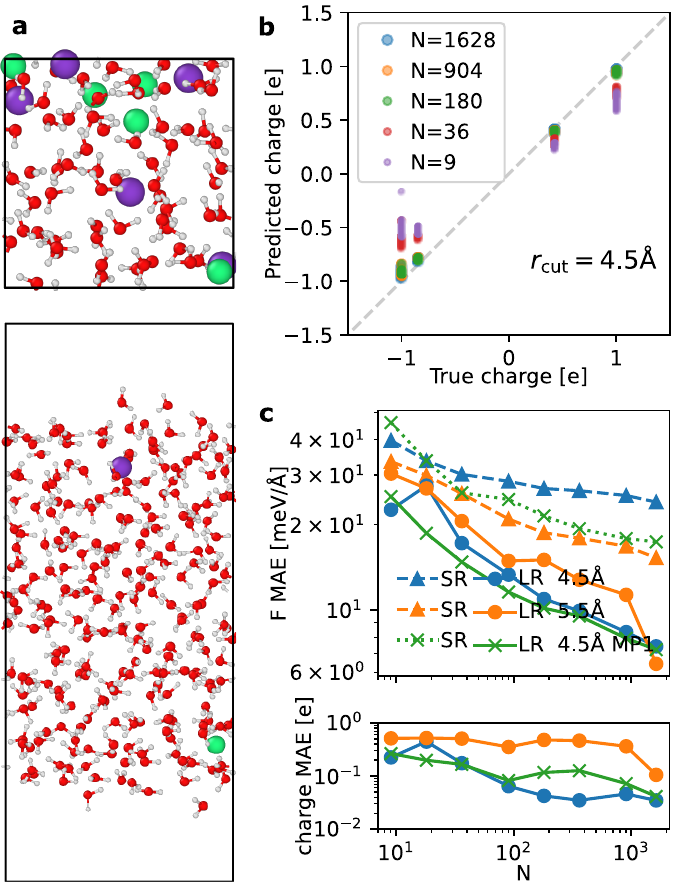}
    \caption{
    \figLabelCapt{a} A bulk electrolyte configuration (upper panel) and an electrolyte-vapor configuration (lower panel).
    \figLabelCapt{b} Comparison of the true and the predicted charges for the CACE-LR models with a cutoff radius of $4.5$~\AA{} and trained on $N$ configurations.
    \figLabelCapt{c} The mean absolute errors (MAEs) on forces, and charges for short-range (SR) and long-range (LR) models trained using different $N$ numbers of samples.
    The MP1 indicates models using one message-passing layer.
    }
    \label{fig:electrolyte}
\end{figure}

We constructed a dataset of potassium fluoride (KF) aqueous solutions with concentrations ranging from 0 to approximately 2 mol/L. 
The dataset includes both bulk electrolyte solution configurations and electrolyte-vapor interfaces, as illustrated in Fig.~\ref{fig:electrolyte}. 
The reference energies and forces were computed using the flexible SPC/Fw water model (with oxygen carrying a charge of $-0.8476$e and hydrogen carrying a charge of $+0.4238$e)~\cite{wu2006flexible}, alongside ions with fixed charges (K: $+1$e, F: $-1$e)~\cite{joung2008determination}.
This electrolyte dataset is significantly more challenging than the random charge example, as it involves multiple species with distinct atomic charges. Additionally, water acts as a dielectric medium, and the presence of interfaces introduces diverse screening effects that vary with depth from the surface.

Fig.~\ref{fig:electrolyte}b shows that the CACE-LR model with $r_\mathrm{cut} = 4.5$~\AA{} is able to recover the true charges after a couple of hundred of training samples.
Fig.~\ref{fig:electrolyte}c shows the learning curves for the MAEs on forces and charges, and the MAEs on energies are all pretty small for all models ($<0.3$~meV/atom for $\gtrapprox$100 samples).
While a larger cutoff or a message passing layer (MP1) improves the SR model,
the LR model with a smaller cutoff $r_\mathrm{cut} = 4.5$~\AA{} achieves better learning efficiency.
Adding a message-passing layer to the LR model has little effect in this case.
This electrolyte example also shows that the LR model is able to learn the charges and energetics of systems involving different species and a dielectric medium that screens electrostatics.

\subsection{Charged molecular dimers}

\begin{figure}
  \centering
\includegraphics[width=0.99\linewidth]{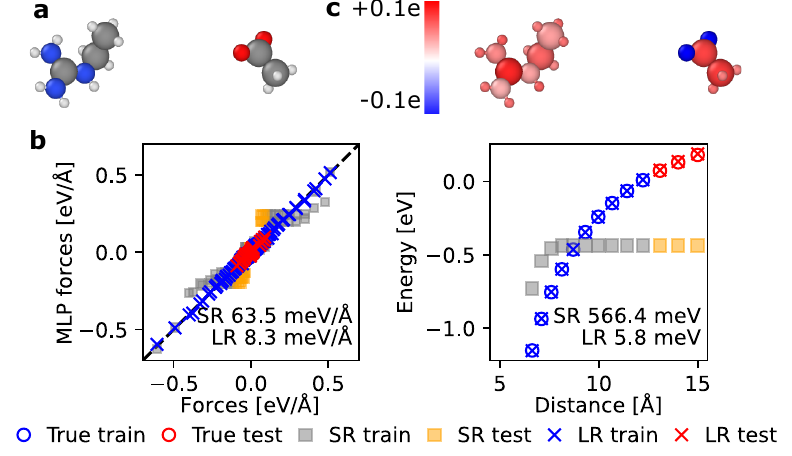}
    \caption{
    \figLabelCapt{a} A snapshot of the molecular dimer configuration of C$_3$N$_3$H$_{10}^+$/C$_2$O$_2$H$_3^-$.
    \figLabelCapt{b} The comparison between the true and predicted force components (left panel), and the binding energy curves (the energy difference between the dimer and two isolated monomers) from SR and LR models.
    \figLabelCapt{c} The predicted charge distribution from the CACE-LR model.
    }
    \label{fig:dimer}
\end{figure}

We revisit an example from a molecular dimer dataset~\cite{burns2017biofragment} used to benchmark LODE~\cite{huguenin2023physics} and LES~\cite{cheng2024latent}. 
This example consists of the binding curve between two charged molecules of C$_3$N$_3$H$_{10}^+$/C$_2$O$_2$H$_3^-$ (shown in Fig.~\ref{fig:dimer}a).
The training set consists of 10 configurations with dimer separation distances between approximately 5~\AA{} and 12~\AA{}, and the test set includes 3 configurations with separations between approximately 12~\AA{} and 15~\AA. 
The dataset includes energy and force information calculated using the HSE06 hybrid DFT with a many-body dispersion correction. 

For the CACE-LR model, here we use a one-dimensional $q$, whereas the original LES paper~\cite{cheng2024latent} used a four-dimensional hidden variable,
and the model test errors are comparable.
Fig.~\ref{fig:dimer}b compares the predicted forces and dispersion curves for the LR and SR models. 
The SR model has one message-passing layer, but as the two molecules can have a distance beyond the cutoff of $r_\mathrm{cut}=5$~\AA, the message-passing scheme does not help.
Fig.~\ref{fig:dimer}c shows the predicted charge distribution.
The total predicted charges on C$_3$N$_3$H$_{10}^+$/C$_2$O$_2$H$_3^-$ molecules are $+1.09$e/$-0.83$e, and $+1.01$e/$-1.01$e after removing the mean charge.
These are very close to the ground truth of $+1$e/$-1$e molecular charges, despite the fact that the MLIP training is agnostic about these charge states.
The two under-coordinated oxygen atoms in C$_2$O$_2$H$_3^-$ have the same strong negative charge, while the rest of the molecule are positively charged.
The undercoordinated carbon in C$_3$N$_3$H$_{10}^+$ has a positive charge, while the other atoms have smaller positive charges.
These trends are consistent with chemical intuitions.

\subsection{Polar dipeptides}

\begin{figure}
  \centering
\includegraphics[width=0.99\linewidth]{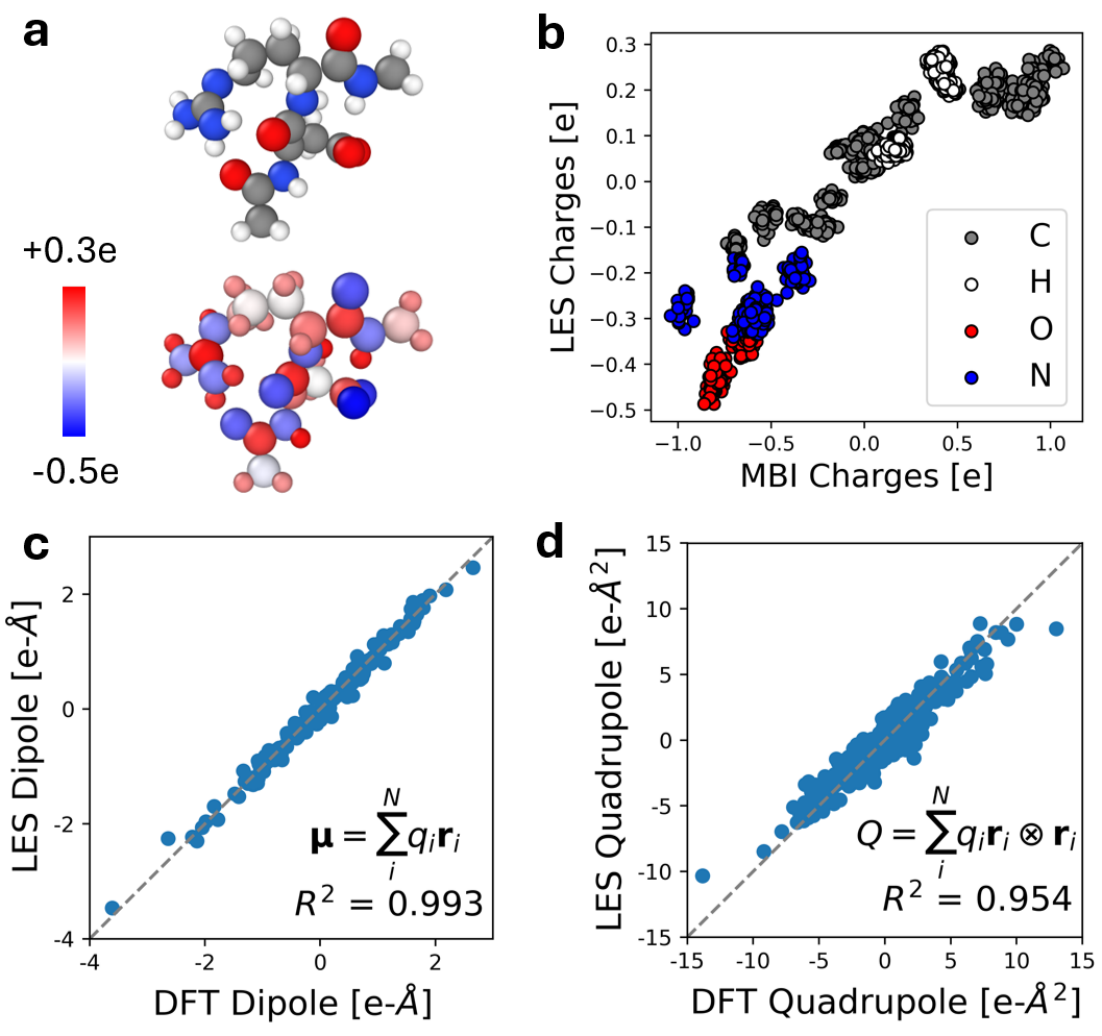}
    \caption{
    \figLabelCapt{a} Top: A snapshot of a dipeptide conformer composed of arginine and aspartic acid from the SPICE dataset~\cite{eastman2023spice}. Bottom: The predicted LES charge distribution.
    \figLabelCapt{b} The predicted charges from LES compared to minimal basis iterative stockholder (MBI) charges in SPICE.
    \figLabelCapt{c} The predicted dipole components computed from the LES charges ($\mathbf{\mu} = \sum_{i=1}^N q_i \mathbf{r}_i$) compared to the DFT dipole components in SPICE.
    \figLabelCapt{d} The predicted traceless quadrupole components computed from the LES charges ($Q = \sum_{i=1}^N q_i \mathbf{r}_i \otimes \mathbf{r}_i)$) compared to the DFT quadrupole components in SPICE.
    }
    \label{fig:dipeptides}
\end{figure}

Since atomic charges in quantum mechanics are not well-defined quantities, a key question is whether the LES charges can be used to predict physical observables such as dipole and quadrupole moments. To answer this question, we turn to the SPICE dataset~\cite{eastman2023spice}, which contains DFT dipole and quadrupole moments as well as minimal basis iterative stockholder (MBI) charges ~\cite{verstraelen2016minimal} for a wide array of drug-like molecules. 
Specifically, we fit CACE-LR on a dataset of polar dipeptides, just by learning from the energy and forces.
Then we determine whether LES is able to infer the DFT dipole and quadrupole moments on a test set of unseen polar dipeptides (illustrated in Fig.~\ref{fig:dipeptides}a). 
We compute the predicted LES dipole via $\mathbf{\mu} = \sum_i^N q_i \mathbf{r}_i$ and quadrupole via $Q = \sum_i^N q_i \mathbf{r}_i \otimes \mathbf{r}_i$ where $q_i$ are the charges predicted by LES and $\mathbf{r}_i$ are the positions of atoms $i$. To make the comparison translationally invariant, we additionally subtract the trace from the calculated and DFT quadrupole moments ($Q' = Q - \frac{1}{3} \mathrm{Tr}(Q) I$). 

Figure \ref{fig:dipeptides}b compares the charges predicted by LES to the MBI charges from SPICE. As is seen, the charges predicted by LES correlate well with the MBI charges, and agree with the usual ordering of electronegativities (O $>$ N $>$ C $>$ H). Furthermore, Fig.~\ref{fig:dipeptides}c compares the dipole moments derived from LES to DFT. Remarkably, we find that the derived dipoles from the LES charges are in excellent agreement with those from DFT ($R^2=0.993$), even though the LR model is not trained explicitly on any charge or dipole information. In absolute terms, the LES mean absolute error (MAE) for dipole moments is 0.089~e-\AA{}, comparable to the 0.063~e-\AA{} MAE of MBI charges derived directly from DFT densities. Fig.~\ref{fig:dipeptides}d further compares the calculated quadrupole moments to those of DFT. Again, we see good agreement of the LES quadrupoles with the physical DFT values ($R^2=0.954$) comparable to MBI ($R^2=0.960$). This remarkable agreement between DFT and LES dipoles and quadrupoles shows that LES is able to convincingly model observables of the molecular charge density even though no charge information is explicitly input into the model training. 

Again, we emphasize that derived atomic charges such as MBI are not physical observables -- although there is significant disagreement between the MBI and LES charges (Fig.~\ref{fig:dipeptides}b), they are both good predictors of the observable molecular dipole and quadrupole moments. The ability of LES to infer dipole and quadrupole moments strongly supports the thesis that MLIPs should be tied to energies and forces rather than any specific definition of atomic charges or electronegativities. 

\subsection{Dataset with different charge states and charge transfer}

\begin{figure}
  \centering
   \includegraphics[width=0.9\linewidth]{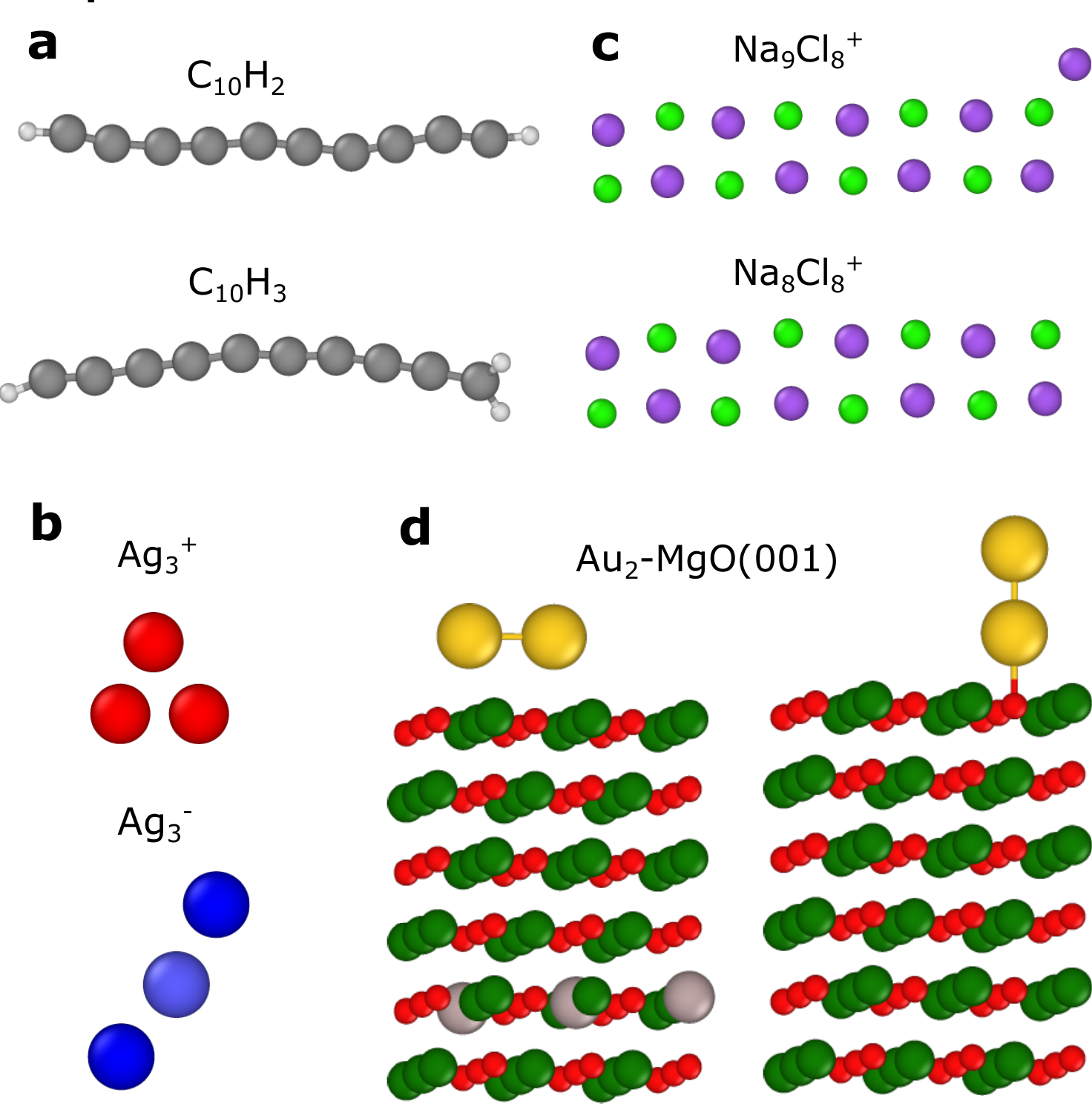}
    \caption{
    Illustrations of the four systems with different charge states and charge transfer, taken from Ref.~\cite{ko2021fourth}.
\figLabelCapt{a} The C$_{10}$H$_{2}$/C$_{10}$H$_{3}^{+}$ set.
\figLabelCapt{b} The Ag$^{+/-}_{3}$ set has Ag trimers in positive or negative charge states.
\figLabelCapt{c} The Na$_{8/9}$Cl$_{8}^{+}$ set.
\figLabelCapt{d} The Au$_{2}$-MgO(001) set has
a wetting (left) or unwetting (right) Au$_{2}$ on the doped (left) or undoped (right) MgO(001) surface.
    }
    \label{fig:systems}
\end{figure}

\citet{ko2021fourth} compiled four datasets 
(C$_{10}$H$_{2}$/C$_{10}$H$_{3}^{+}$,
Ag$^{+/-}_{3}$, 
Na$_{8/9}$Cl$_{8}^{+}$,
and Au$_{2}$ on MgO(001),
illustrated in Fig.~\ref{fig:systems}) 
that specifically target systems in different charge states or where charge transfer mediated by long-range electrostatic interactions is significant.
In Table~\ref{table:1} we compare the CACE-LR errors with the values obtained with 3G-HDNNP and 4G-HDNNP~\cite{ko2021fourth}, as well as a charge constraint ACE model through a local many-body expansion ($\chi$+$\eta$(ACE))~\cite{rinaldi2024charge}. 
The comparison between CACE and ACE is a rather direct one: their descriptors are mathematically equilvalent~\cite{dusson2022atomic}.
4G-HDNNP and $\chi$+$\eta$(ACE) both fit charges explicitly,
while CACE-LR only fits to energy and forces and no total charge constraint was used.
We used a 90\% train and 10\% test split, consistent with Ref.~\cite{ko2021fourth}.

\begin{table*}[t]
	\setlength{\tabcolsep}{6pt} %
    \renewcommand{\arraystretch}{1.0}
\centering
\begin{tabular}{|c | c | c | c | c | c| c |}
\hline
             & & ACE  & $\chi$ + $\eta$(ACE) & 3G-HDNNP & 4G-HDNNP & CACE-LR\\
\hline
     & $r_\mathrm{cut}$ & 6~\AA & 6~\AA & 4.23~\AA & 4.23~\AA & 4.23~\AA{}\\
     C$_{10}$H$_{2}$/C$_{10}$H$_{3}^{+}$ & E & 0.76  & 0.75  & 2.045 & 1.194 & 0.73\\
     & F & 37.22 & 35.16 & 231.0 & 78.00 & 36.9\\
\hline
     & $r_\mathrm{cut}$ & 6~\AA & 6~\AA & 5.29~\AA & 5.29~\AA  & 5.29~\AA\\
    Ag$^{+/-}_{3}$ & E  & 809.62 & 0.21  & 320.2 & 1.323 & 0.162 \\
     & F & 285.81 & 23.10 & 1913  & 31.69 & 29.0 \\
\hline
    & $r_\mathrm{cut}$ & 6~\AA & 6~\AA & 5.29~\AA & 5.29~\AA & 5.29~\AA{}\\
    Na$_{8/9}$Cl$_{8}^{+}$ & E  & 1.55   & 0.71  & 2.042 & 0.481  & 0.21\\
    & F  & 41.72  & 12.35 & 76.67 & 32.78 & 9.78\\
\hline
    & $r_\mathrm{cut}$ & 6~\AA & 6~\AA & - & 4.23~\AA & 5.5~\AA \\
    Au$_{2}$-MgO(001) & E & 2.56   & 1.63  & -  & 0.219 & 0.073 \\
    & F & 88.70  & 50.27 & -  & 66.00 & 7.91 \\
\hline
\end{tabular}
    \caption{Test root mean squared errors (RMSE) are reported for energies in meV/atom, forces in meV/\AA. The 3G-HDNNP and 4G-HDNNP values are from~\cite{ko2021fourth}.
    The ACE and $\chi$ + $\eta$(ACE) values are from ~\cite{rinaldi2024charge}.
    Short-ranged CACE with embedded charge states was used for the Ag$^{+/-}_{3}$ system.
    }
    \label{table:1}
\end{table*}

The C$_{10}$H$_{2}$/C$_{10}$H$_{3}^{+}$ set contains
carbon chains terminated with hydrogen atoms in the neutral or positively charged state.
With and without the added proton on the right-hand side of Fig.~\ref{fig:systems}a, the atoms in the left half of the molecule can have almost identical environments but different atomic charges, which results in high fitting errors in 3G-HDNNP~\cite{ko2021fourth} due to the contradictory information. 

The Ag$^{+/-}_{3}$ example illustrated in Fig.~\ref{fig:systems}b contains
Ag trimers in two different charge states.
As the system size is small such that there are no long-range interactions, 
we used only short-ranged CACE MLIP with embedded charge states.
Since the energies depend on the overall charge states of the clusters,
this causes the degeneracy issue between atomic structures and potential energy surfaces,
leading to the poor performance of the 3G-HDNNP and the charge-agnostic ACE methods.
Both the charge constraint $\chi$+$\eta$(ACE) model and the charge-state-embedded CACE lift such degeneracies, leading to drastically improved descriptions.

The Na$_{8/9}$Cl$_{8}^{+}$ set (Fig.~\ref{fig:systems}c) contains the ionic Na$_{9}$Cl$_{8}^{+}$ clusters and Na$_{8}$Cl$_{8}^{+}$ when a neutral Na atom is removed.
This is also an example where global charge transfer is present.
CACE-LR achieves the lowest errors in this case.

The Au$_{2}$-MgO(001) set (Fig.~\ref{fig:systems}d has
a diatomic gold cluster supported on the MgO(001) surface with two adsorption
geometries: an upright non-wetting orientation of the dimer attached to a surface oxygen, and a parallel
wetting configuration on top of two Mg atoms. 
Moreover, three Al dopant atoms were introduced into the fifth layer below the surface (the gray atoms in the left panel of Fig.~\ref{fig:systems}d).
Despite having large distances of more than 10 \AA,
the dopant atoms have a major influence on the electronic structure and the relative stability between the wetting and the non-wetting configurations. 

\begin{table}[h!]
\centering
\begin{tabular}{|c|c|c|c|c|c|}
\hline
 & DFT & 2G-HDNNP &   CACE-SR & 4G-HDNNP & CACE-LR 
\\ \hline
Doped     & -66.9               & 375        & 431       & -41    & -70.6                \\ 
Undoped   & 934.8               & 375       & 431        & 975    & 931.3                 \\ \hline
\end{tabular}
\caption{Energy difference ($E_\mathrm{wetting}-E_\mathrm{nonwetting}$) in meV between the wetting and nonwetting configurations for doped and undoped substrates.}
\label{tab:e_diff}
\end{table}

\begin{figure}
  \centering
   \includegraphics[width=0.9\linewidth]{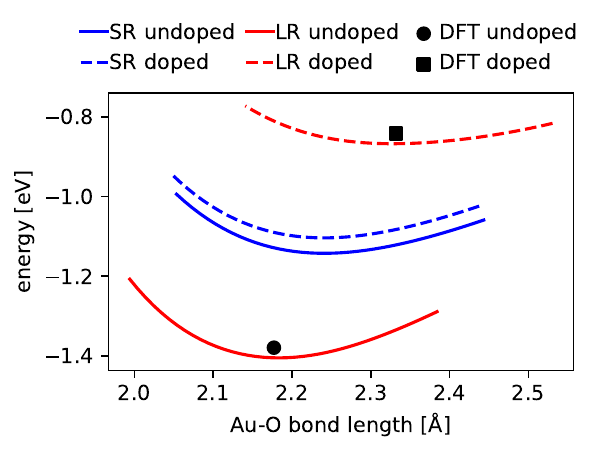}
    \caption{
Potential energies for the Au$_2$ cluster adsorbed at the MgO(001) substrate for the non-wetting geometry for the Al-doped and undoped cases. The equilibrium DFT bond lengths DFT energy and the associated minimum energies are denoted in black symbols. 
The Au–O bond length is the minimum distance between Au and O atoms.
}
    \label{fig:au-pes}
\end{figure}

In this example, CACE-LR achieves errors that are approximately an order of magnitude smaller than those of the other methods compared.
As an additional test, we performed geometry optimizations of the positions of the gold atoms, with the substrate fixed, for both doped and undoped surfaces. The results were compared to reference DFT calculations and previous results using the 4G-HDNNP method~\cite{ko2021fourth}. For the pure MgO substrate, the non-wetting configuration is energetically favored, whereas doping stabilizes the wetting geometry. The energy differences between the wetting and non-wetting configurations for both doped and undoped substrates are presented in Table~\ref{tab:e_diff}.
Short-range models, such as 2G-HDNNP and CACE-SR, predict nearly degenerate energy values for these configurations, as expected. In contrast, CACE-LR delivers highly accurate predictions, closely matching the reference results.
Consistent with findings in \cite{ko2021fourth}, we also present the potential energy surface (PES) for the non-wetting geometry on doped and undoped substrates as a function of the distance between the bottom Au atom and its neighboring oxygen atom, shown in Fig.~\ref{fig:au-pes}. Equilibrium bond lengths and energies derived from DFT are marked with black symbols. Notably, CACE-LR accurately resolves the distinct equilibrium bond lengths, with a slight shift in the potential energy surface likely attributable to differences in DFT convergence settings.

\begin{figure}
  \centering
   \includegraphics[width=0.75\linewidth]{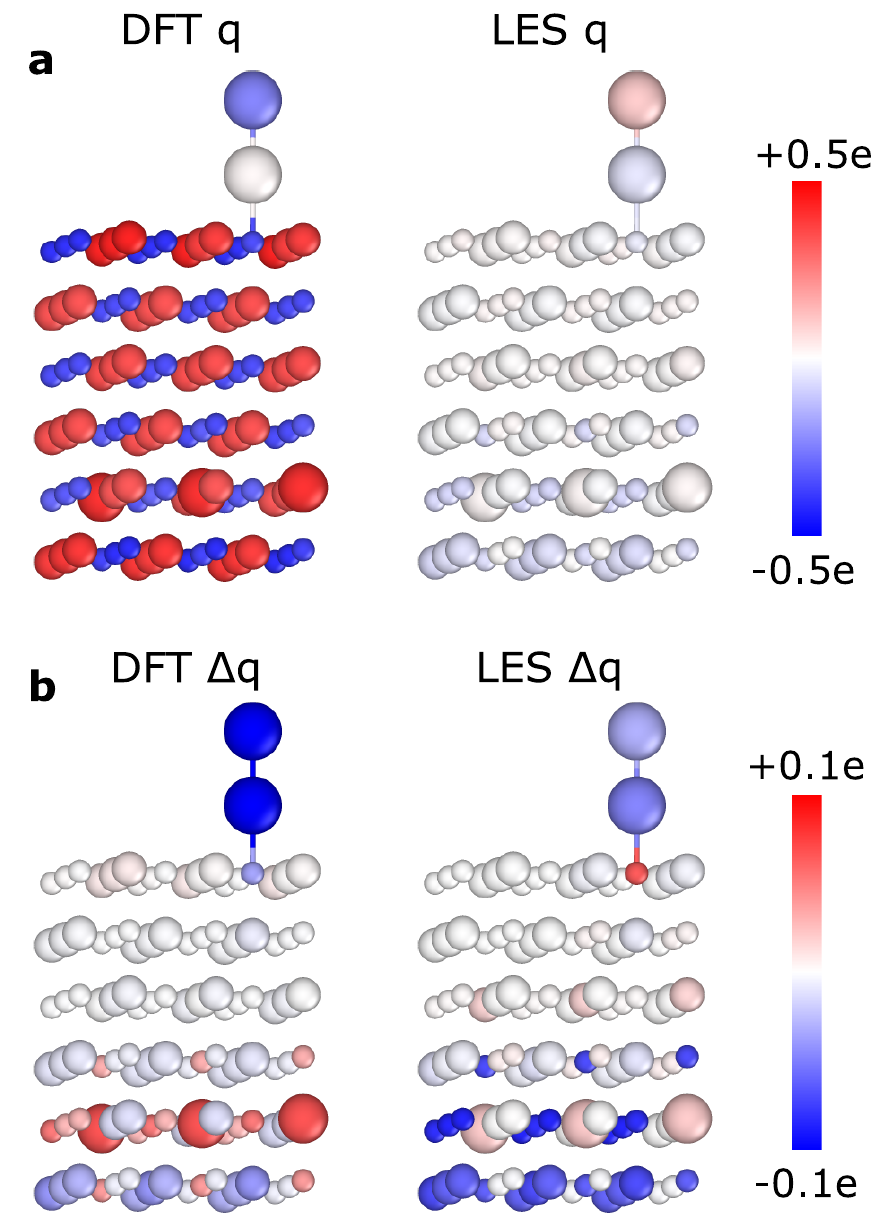}
    \caption{
\figLabelCapt{a} The atomic charges from the underlying DFT data (left), and the predicted atomic charge from CACE-LR (right) for the nonwetting Au$_2$ cluster adsorbed on the doped MgO(001) substrate.
\figLabelCapt{b} The change of atomic charges due to dopping, from the DFT data (left), and the predicted atomic charge from CACE-LR (right).
    }
    \label{fig:au-q}
\end{figure}

We rationalize why the CACE-LR method delivers significantly more accurate predictions compared to other long-range methods that explicitly fit atomic charges.
In Fig.~\ref{fig:au-q}a, we compare the atomic charge distribution from the underlying DFT data, obtained via Hirshfeld population analysis~\cite{ko2021fourth,hirshfeld1977bonded}, with the charges predicted by CACE-LR. The charges from CACE-LR are generally much smaller in magnitude and are primarily localized on the Au dimer and the dopant. In contrast, the DFT charges show sharp positive values for metal atoms and sharp negative values for oxygen atoms in the substrate.
We hypothesize that explicitly modeling such DFT-derived charges for metals and oxygen is unnecessary for accurately predicting energy and forces. Short-ranged MLIPs are already well-suited to describe bulk oxides without dopants due to the screening effects that diminish the influence of these charge extremes.
In Fig.~\ref{fig:au-q}b, we plot the changes in atomic charges resulting from doping, by taking the atomic charge difference for each atom from relaxed doped and doped structures, which shows a clear correlation between DFT and CACE-LR results. This example suggests that the charges predicted by CACE-LR can be interpreted as response charges rather than DFT partial charges, focusing on the aspects of charge redistribution relevant to energy and force predictions.

\subsection{Electrolyte/solid interfaces}
As example applications to electrolyte/solid interfaces, we selected two sets of systems.
The first is the Pt(111)/KF(aq) interface dataset from Ref.~\cite{zhu2024machine}, which describes the Pt electrode with the (111) surface forming an interface with K and F ions in water solutions.
For training the MLIP,
Ref.~\cite{zhu2024machine} used a DPLR model:
the short-ranged part is a standard Deep Potential (DP) model with a cutoff of 5.5~\AA,
and the long-range electrostatics is computed using spherical Gaussian charges associated with the nuclei (i.e., 6e, 1e, 9e, 7e, and 0e for O, H, K, F and Pt atoms, respectively) and the
average positions of the MLWCs~\cite{zhang2022deep} with a total charge of $-8$e  associated with each O, K, and F atoms.
Note that such MLWC schemes are not applicable to conductors, so Ref.~\cite{zhu2024machine} used the classical Siepmann-Sprik model~\cite{siepmann1995influence} to describe the Pt electrode in MD simulations.

The second dataset from Ref.~\cite{zhang2024electrical} is for modeling 
the anatase TiO$_2$ (101) surface in contact with NaCl-water electrolyte
solutions at various pHs.
This dataset comprehensively spans the configurational space of bulk anatase TiO$_2$,
water, and various aqueous electrolyte solutions (NaCl, NaOH, HCl, and their mixtures), as well as anatase (101) interfaces with each of these liquids.
Ref.~\cite{zhang2024electrical} trained a standard short-ranged DP and a DPLR MLIP.
The LR part in the DPLR model is also based on the electrostatics of spherical Gaussian charges associated with the ions (nuclei + core electrons) and the valence electrons.
More specifically, 4e, 1e, 6e, 9e, and 7e for Ti, H, O, Na, and Cl ions, and each O, Na, and Cl ion has four WCs each carrying -2e.

We fitted the CACE-SR and CACE-LR models, without message passing.
The results are presented in Table~\ref{table:error-interface}.
We speculate that the improved performance of the CACE models compared to the DP models can be attributed to two reasons:
First, the DP descriptors are restricted to two-body and three-body terms,
while the ACE framework can include higher-body-order interactions and in this case we truncate to four-body terms.
The inclusion of higher-body terms makes the model more expressive and helps alleviate the degeneracy problem~\cite{pozdnyakov2020incompleteness}.
Second, the LES scheme allows each atom to carry a flexible, learned, charge, in contrast with the fixed charge in the DPLR method.

\begin{figure}
  \centering
  \includegraphics[width=0.99\linewidth]{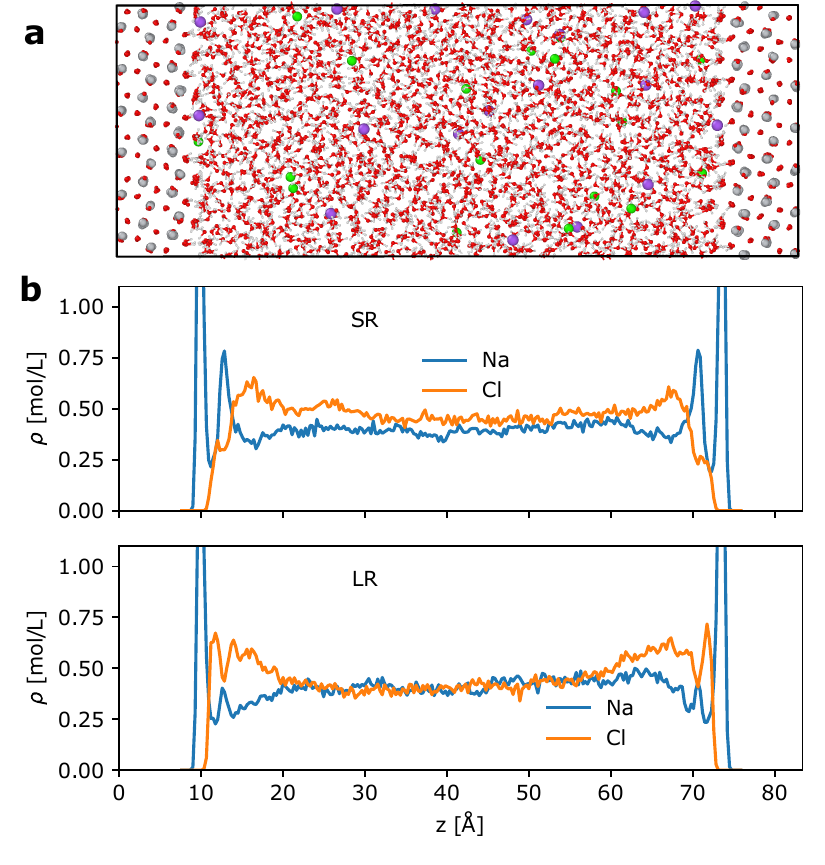}
    \caption{
    \figLabelCapt{a} The system of the anatase TiO$_2$ (101) surface and NaCl in water solution.
    \figLabelCapt{b} Plane-averaged ion distributions along the $z$-direction for the TiO$_2$-NaCl solution interface obtained from CACE-SR and CACE-LR MD simulations at 600~K. 
    }
    \label{fig:tio2}
\end{figure}

To showcase the effect of long-range interactions on the structures of the electrolyte and the electric double layer (EDL), we performed
MD simulations at 600~K for 5~ns on a large system of anatase TiO$_2$ surface and NaCl in water solution (illustrated in Fig.~\ref{fig:tio2}).
This is also a test that was performed in Ref.~\cite{zhang2024electrical}.
Fig.~\ref{fig:tio2} shows the ion distributions obtained from the MD simulations using the CACE-SR and CACE-LR models. 
In reality, the solution should recover its bulk properties in the central region that is away from the interface and have equal densities of Na$^+$ and Cl$^-$ ions. 
However, the SR model, lacking long-range electrostatic interactions, imposes no energy penalty for unphysical charge imbalances. Consequently, the MD simulation predicts an excess Cl$^-$ density of approximately 0.05 mol/L in the center of the box.
In contrast, incorporating long-range interactions with the CACE-LR model eliminates this artifact and alters the ion distributions within the EDL. 
These effects, including the correction of charge imbalance and modified EDL structures, were also reported in Ref.~\cite{zhang2024electrical}. Notably, the CACE-LR model predicts a significantly lower second Na$^+$ density peak near the interface compared to Ref.~\cite{zhang2024electrical}.

\begin{table*}[t]
	\setlength{\tabcolsep}{6pt} %
    \renewcommand{\arraystretch}{1.0}
\centering
\begin{tabular}{|c | c | c | c | c | c| }
\hline
             & & DPSR  & DPLR & CACE-SR & CACE-LR\\
\hline
     & $r_\mathrm{cut}$ & - & 5.5~\AA & 5.5~\AA & 5.5~\AA \\
     Pt(111)/KF(aq) & E & -   & 1.305  & 0.863 & 0.309 \\
     & F & -   & 75.00  & 58.6 & 34.1 \\
\hline
     & $r_\mathrm{cut}$ & 6~\AA & 6~\AA & 5.5~\AA & 5.5~\AA \\
     TiO$_2$(101)/NaCl+NaOH+HCl(aq) & E & 0.88  & 0.79 & 0.721 & 0.435 \\
     & F & 124   & 119  & 103 & 70.5 \\
\hline
\end{tabular}
    \caption{Test root mean squared errors (RMSE) are reported for energies in meV/atom, forces in meV/\AA. 
    The DPLR results for the Pt(111)/KF(aq) set are from Ref.~\cite{zhu2024machine},
    and DPSR and DPLR results for the TiO$_2$(101)/NaCl+NaOH+HCl(aq) set are from Ref.~\cite{zhang2024electrical},
    }
    \label{table:error-interface}
\end{table*}

\subsection{Solid-solid interface}

\begin{figure}
  \centering
\includegraphics[width=\linewidth]{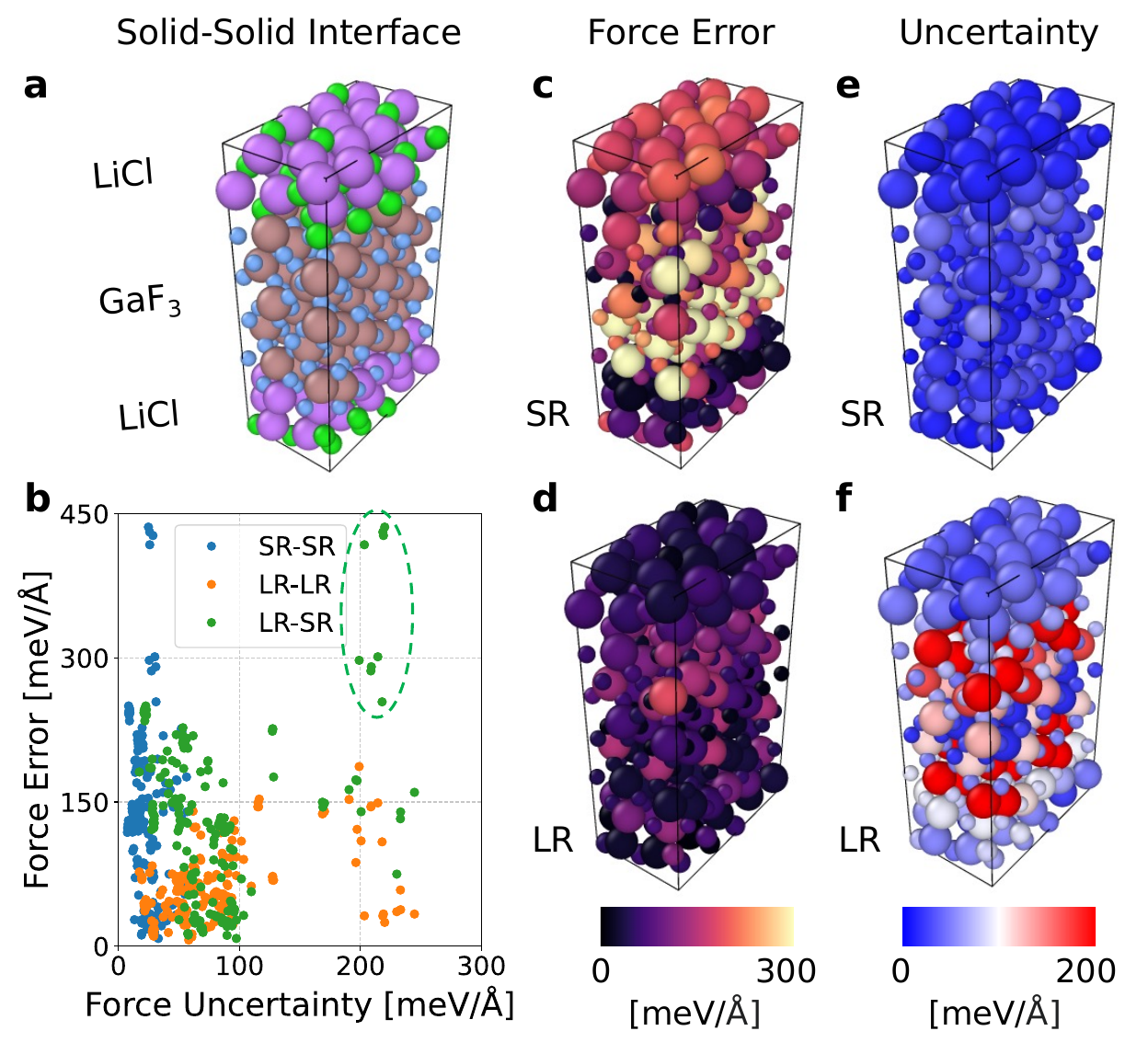}
\caption{\figLabelCapt{a} A DFT-relaxed structure of the LiCl(001)/GaF$_3$(001) interface. \figLabelCapt{b} Correlation between force errors and uncertainties computed from ensemble predictions. Blue: SR model error vs. SR model uncertainty; Orange: LR model error vs. LR model uncertainty; Green: SR model error vs. LR model uncertainty. \figLabelCapt{c}-\figLabelCapt{f} Atomic-resolved force errors (left panels) and uncertainty estimates (right panels) for SR (top) and LR (bottom) models.}
\label{fig:solid-solid}
\end{figure}

Atomistic modeling of solid-solid interfaces is essential in understanding material synthesizability \cite{Miura2021}. The heterogeneous nature of these interfaces requires long-period structures, particularly in cases involving charge transfer, which necessitates long-range descriptions beyond conventional MLIPs. To evaluate the predictive accuracy of our models, we conducted a benchmark study comparing CACE-SR and CACE-LR using the LiCl(001)/GaF$_3$(001) interfacial system \cite{Gupta2023}. The training dataset includes bulk and interfacial configurations in the LiCl-GaF$_3$ chemical space with corresponding DFT-calculated energies and interatomic forces. To assess model uncertainty, we trained an ensemble of four SR/LR models and used their predictions to estimate force uncertainties (see Methods).
For in-distribution (ID) test set performance, CACE-SR and CACE-LR models achieve RMSEs of 78.8 meV/\AA\ and 67.8 meV/\AA, respectively. 

To evaluate model transferability, we constructed an out-of-distribution (OOD) test set using a large solid-solid heterostructure relaxed with DFT calculations ($\sim$30 \AA\ in the $z$-direction, Fig.~\ref{fig:solid-solid}a). This extended structure, containing eight Ga layers and four Li layers, represents a more realistic interface with much reduced finite-size effects compared to the training configurations. 
On this OOD set, the LR model demonstrates improved predictive accuracy with a force component error of 40.5 meV/\AA\ compared to 116.3 meV/\AA\ for the SR model. The atomic-resolved force errors are visualized in Figs. \ref{fig:solid-solid}c and \ref{fig:solid-solid}d, which were computed from the square root of the sum of force component errors in $x,y,z$-directions.

Force uncertainties were quantified using ensemble variance from the four trained models. The SR model exhibits lower uncertainties (Fig.~\ref{fig:solid-solid}e), indicating a better parametrization on the ID training set. In contrast, the LR model shows elevated uncertainties (Fig.~\ref{fig:solid-solid}g), effectively identifying OOD atomic environments in the heterostructure. The correlation between the absolute force errors (RMSE against DFT) and uncertainties is shown in Fig.~\ref{fig:solid-solid}b, where green dots specifically highlight the relationship between SR model errors (poor prediction) and LR model uncertainties (OOD detection). Interestingly, the LR model identifies regions of SR model failure (green dashed circle in Fig.~\ref{fig:solid-solid}b), which are further evidenced by the spatial correspondence in Figs.~\ref{fig:solid-solid}c and \ref{fig:solid-solid}f. 
These results suggest that despite the SR MLIPs achieving adequate ID performance for this system, they lack the mathematical framework to capture long-period structure features that are essential for electrostatic interactions. In contrast, the LR models with LES overcome this limitation with improved transferability. More generally, the enhanced OOD detection capabilities are essential for robust uncertainty quantification in broader applications such as materials property predictions and generation \cite{dai2024uncertainty}. While our current implementation relies on computationally intensive ensemble variance, the LES framework is compatible with various uncertainty quantification methods, including Gaussian mixture models \cite{zhu2023fast}, Monte Carlo dropout \cite{gal2016_MC_dropout}, and deep evidential regression \cite{amini2020deep}.

\section{Discussion}

The LES framework is highly interpretable in physical terms: the hidden variable $q$, when restricted to one dimension for computing electrostatic long-range potentials, corresponds to the partial charges on atoms.
In cases such as random charges and electrolyte solutions, where the underlying potential energy surfaces are described by classical forcefields with fixed charges, LES accurately recovers those charges. For quantum-mechanical systems, such as those described using DFT, the LES-derived partial charges can be understood as a coarse-grained approximation of the net electrostatic effect of electron density polarization. This approximation has also been rationalized and applied to parameterize scaled charges in classical forcefields~\cite{kirby2019charge}.

Notably, atomic charges in quantum-mechanical systems are not physical observables. 
In DFT, there exists a wide variety of methods to assign local atomic charges given the global charge density, each providing different frameworks and values~\cite{wiberg1993comparison,marenich2012charge}. These include Mulliken population analysis that relies on the overlap of atomic orbitals~\cite{mulliken1955electronic}, shareholder methods such as Hirshfeld population analysis~\cite{hirshfeld1977bonded} and MBI~\cite{verstraelen2016minimal}, Born effective charges derived from perturbation theory~\cite{gonze1997dynamical}, and parameterizations based on these schemes \cite{marenich2012charge}.
In the case of charge dimers and polar dipeptides, LES-derived charges are consistent with chemical intuition. For polar dipeptides, as shown in Fig.~\ref{fig:dipeptides}, LES charges are correlated with, but not equivalent to, MBI charges. Yet, despite this imperfect correlation, LES charges reproduce DFT dipoles and quadrupoles with remarkable accuracy ($R^2 = 0.993$ and $R^2 = 0.954$). This ``free lunch''--predicting dipole and quadrupole moments without explicitly learning the multipoles or charges--highlights the physical interpretability embedded in the LES framework.

Indeed, the ambiguity of DFT-assigned charges suggests that directly learning such charges may not be necessary for -- or may even be a detriment to -- constructing accurate interatomic potentials. 
This insight is supported by results for four challenging systems involving different charge states and charge transfer (C$_{10}$H$_{2}$/C$_{10}$H$_{3}^{+}$, Ag$^{+/-}_{3}$, Na$_{8/9}$Cl$_{8}^{+}$, and Au$_{2}$ on MgO(001)), where CACE-LR outperformed both 4G-HDNNP~\cite{ko2021fourth} and $\chi$ + $\eta$(ACE)\cite{rinaldi2024charge}, which explicitly learn charges and perform charge equilibration (see Table\ref{table:1}).
For interfacial systems, such as Pt(111)/KF(aq) and TiO$_2$(101)/NaCl+NaOH+HCl(aq), CACE-LR also achieved greater accuracy compared to DPLR, which learns the positions of Wannier centers (see Table~\ref{table:error-interface}).

One can further speculate that the improved performance of LES compared to the other methods stems from the fact that LES does not directly learn from charges. For instance, in the Au$_2$-MgO system, LES achieves an error an order of magnitude lower than 4G-HDNNP~\cite{ko2021fourth} and $\chi$ + $\eta$(ACE)~\cite{rinaldi2024charge}. This likely results from LES capturing the response charge—changes in atomic charges due to doping—rather than the sharply peaked and method-dependent DFT charges, as illustrated in Fig.~\ref{fig:au-q}.
While our test uses simple metal oxides, the response charge formalism is particularly relevant for complex ionic systems, such as transition metal oxides. Previous studies have shown that materials with localized $d$-electrons exhibit self-regulating response in DFT~\cite{Wolverton1998_LiCoO2}, where the system maintains constant local charge on transition metal atoms by minimizing external perturbations through rehybridization~\cite{raebiger2008charge}. Given this complexity and the fact that DFT charges vary depending on the computation method, directly inferring them introduces inefficiencies in resolving their ambiguous components~\cite{Walsh2018}. The strong performance of LES suggests that the detailed prediction of atomic charges is less critical; instead, the primary focus should remain on accurately predicting physically observable quantities, such as energies and forces.
Moreover, by avoiding the direct learning of charges, LES circumvents the need for explicit charge equilibration, thereby reducing the associated computational overhead.

Omitting long-range interactions can result in severely inaccurate predictions for many systems. For example, standard short-ranged MLIPs fail to predict the binding curve of a charged molecular dimer (Fig.~\ref{fig:dimer}), cannot distinguish the different adsorption behaviors of Au dimers on doped and undoped MgO substrates (Table~\ref{tab:e_diff} and Fig.~\ref{fig:au-pes}), and even produce a charge imbalance in the bulk region of the TiO$_2$-NaCl(aq) solution interface (Fig.~\ref{fig:tio2}).
Alarmingly, the commonly used ensemble uncertainty quantification method was unable to detect the large errors of SR MLIPs in out-of-distribution cases, such as the solid-solid LiCl(001)/GaF$_3$(001) interface. 
This highlights that standard SR models can yield unphysical results in certain systems, and these errors may go unnoticed when relying solely on conventional uncertainty quantification techniques.

In summary, we show that LES is a physics-informed approach that enables learning of atomic charges and long-range interactions directly from energies and forces, without requiring explicit charge labels or additional input. 
The framework consistently provides superior accuracy in modeling long-range interactions compared to existing MLIPs.
We thus demonstrate LES to be a versatile and efficient tool for addressing a wide range of challenging systems where long-range interactions play a critical role, such as electrolyte interfaces, charged molecular complexes, and ionic solutions.

\section{Methods}

\paragraph{Random charges}
The dataset contains a total of 1000 configurations, and each configuration has 64 atoms with +1e charge and 64 atoms with -e charge.
The set was collected from NPT simulations at 4000~K and zero external pressure.
We performed the NPT simulations and computed the energy and forces in LAMMPS, using the Nose-Hoover thermostat and barostat.
The standard deviations in energy and forces are 0.17~eV/atom and 2.0~eV/\AA, respectively.

For the CACE representation, we used 6 Bessel radial functions with $c = 12$, $l_\mathrm{max} = 3$, $\nu_\mathrm{max} = 3$, $N_\mathrm{embedding} = 3$, no message passing, and different cutoff of $r_\mathrm{cut} = 4.5$~\AA, $5$~\AA, or $5.5$~\AA.
For the long-range component, we used a 1-dimensional $q$, $\sigma = 1$~\AA{}, and a maximum cutoff of $k_c = 2\pi$ ($dl=1$~\AA{} in the CACE LES syntax) in the Ewald summation.

\paragraph{Electrolyte solution}
The dataset of KF aqueous solution 
contains both bulk electrolyte solution configurations (1206 configurations with 64 water molecules and 0-5 ion pairs), and electrolyte-vapor interfaces (603 configurations with 225 water molecules and 1, 2, or 10 ion pairs).
We performed NVT MD simulations at 370~K to collect snapshots using the Nose-Hoover thermostat in LAMMPS,
employing SPC/Fw water~\cite{wu2006flexible} (O has charge -0.8476e, H has charge +0.4238e), and ions with fixed charges (K has charge +1e, F has charge -1e) and Lennard-Jones interactions~\cite{joung2008determination}.
The standard deviations in energy and forces are 0.074~eV/atom and 0.9~eV/\AA, respectively.

For the CACE representation, we used 6 Bessel radial functions with $c = 12$, $l_\mathrm{max} = 3$, $\nu_\mathrm{max} = 3$, $N_\mathrm{embedding} = 4$, no message passing ($T=0$) or one message passing layer ($T=1$), and different cutoff of $r_\mathrm{cut} = 4.5$~\AA, or $5.5$~\AA.
For the long-range component, we used a 1-dimensional $q$, $\sigma = 1$~\AA{}, and a maximum cutoff of $k_c = \pi$ ($dl=2$~\AA) in the Ewald summation.

\paragraph{Polar dipeptides}
The dataset of polar dipeptides was taken from the SPICE dataset developed by \citet{eastman2023spice}. The dataset contains energies and forces, for a large number of drug-like molecules including a complete set of dipeptides formed from 26 amino acid variations. The subset used in Figure \ref{fig:dipeptides} consists of dipeptides with one positively charged amino acid (arg, lys, or hip) and one negatively charged amino acid (glu or asp), resulting in a total of 12 dipeptides (with both ways of bonding together two amino acids included, e.g. glu-arg or arg-glu) with 50 conformers each. We retain the conformers of one of the 12 dipeptides as a test set and 10\% of the remaining structures as a validation set.

\begin{table}[h!]
\centering
\begin{tabular}{|l|c|c|c|c|}
\hline
{} &  CACE-SR &  CACE-LR &  CACE-SR &  CACE-LR \\
{} &  Val & Val & Test & Test \\
\hline
E &         1.72 &         1.45 &          2.35 &          1.89 \\
\hline
F &        58.07 &        52.18 &         72.43 &         61.13 \\
\hline
\end{tabular}
\caption{Performance of CACE-SR and CACE-LR on the validation and test sets of the 12 polar dipeptides. Errors are reported via RMSE in meV/atom for energy and in meV/\AA{} for forces.}
\label{tab:dipep}
\end{table}

Table ~\ref{tab:dipep} shows the RMSE performance of both CACE-LR and CACE-SR in determining the energies and forces of these dipeptides. CACE-LR provides slightly better forces and errors than CACE-SR as well as better generalizability to the conformers of the unseen dipeptide (glu-arg). For the CACE model, we used $r_{cut}=4.0$ \AA{}, 6 trainable Bessel radial functions, $c = 12$, $l_\mathrm{max} = 4$, $\nu_\mathrm{max} = 3$, one message passing layer ($T=1$), and different embeddings of sender and receiver nodes with $N_\mathrm{embedding} = 4$. For LES, we used $\sigma=1.5$ \AA{} and the long-range energy from Eq.~\eqref{eq:e_tot} was computed in real space as the configurations are with aperiodic conditions.

\begin{figure}
  \centering
\includegraphics[width=0.99\linewidth]{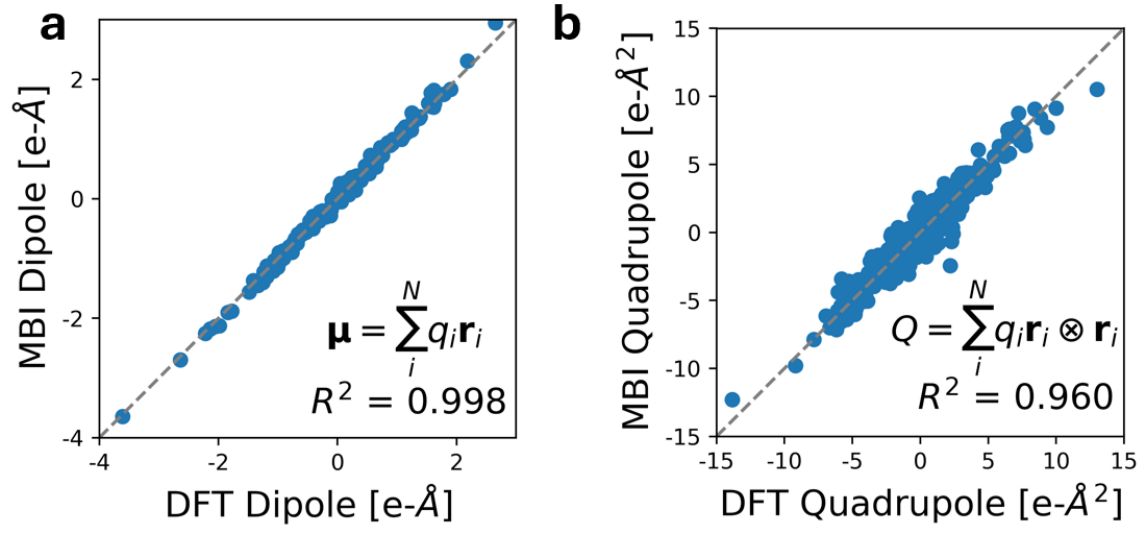}
    \caption{
    \figLabelCapt{a} The predicted dipole components computed from the MBI charges ($\mathbf{\mu} = \sum_{i=1}^N q_i \mathbf{r}_i$) compared to the DFT dipole components in SPICE.
    \figLabelCapt{b} The predicted traceless quadrupole components computed from the MBI charges ($Q = \sum_{i=1}^N q_i \mathbf{r}_i \otimes \mathbf{r}_i)$) compared to the DFT quadrupole components in SPICE.
    }
    \label{fig:mbi-dipeptides}
\end{figure}

Fig.~\ref{fig:mbi-dipeptides} shows the comparison of dipoles and quadrupoles calculated with MBI charges (derived from the DFT density) to the DFT dipoles and quadrupoles. As is seen, the results are very similar to dipoles and quadrupoles calculated with the LES charges (Fig. ~\ref{fig:dipeptides}).

\paragraph{Charged molecular dimers}

The LODE molecular dimer dataset includes energy and force information calculated using the HSE06 hybrid density functional theory (DFT) with a many-body dispersion correction. We used the molecular pair with id 0.

The CACE representation uses
a cutoff $r_\mathrm{cut} = 5$~\AA, 6 Bessel radial functions, $c = 8$, $l_\mathrm{max} = 2$, $\nu_\mathrm{max} = 2$, $N_\mathrm{embedding} = 3$, and one message passing layer ($T = 1$). 
The long-range component $E^{lr}$ employed a 1-dimensional hidden variable computed from the same CACE $B$-features and utilized Ewald summation with $\sigma = 1$~\AA{} and a $\mathbf{k}$-point cutoff of $k_c = 2\pi/3$ ($dl=3$~\AA). 

\paragraph{4G-HDNNP dataset}
The four datasets 
(C$_{10}$H$_{2}$/C$_{10}$H$_{3}^{+}$,
Ag$^{+/-}_{3}$, 
Na$_{8/9}$Cl$_{8}^{+}$,
and Au$_{2}$-MgO(001)) are from Ref.~\cite{ko2021fourth}.

For C$_{10}$H$_{2}$/C$_{10}$H$_{3}^{+}$,
we used $r_\mathrm{cut} = 4.23$~\AA{} (8 Bohr) which is the same as the cutoff in Ref.~\cite{ko2021fourth}, 6 Bessel radial functions, $c = 8$, $l_\mathrm{max} = 3$, $\nu_\mathrm{max} = 3$, $N_\mathrm{embedding} = 2$, no message passing,
1-dimensional hidden variable, $\sigma = 1$~\AA{}, and $k_c = \pi$ ($dl=2$~\AA). 

For Ag$^{+/-}_{3}$,
we used $r_\mathrm{cut} = 5.29$~\AA{} (10 Bohr), 6 Bessel radial functions, $c = 8$, $l_\mathrm{max} = 3$, $\nu_\mathrm{max} = 3$, $N_\mathrm{embedding} = 1$, no message passing,
total charge state embedding, and no long-range component. 

For Na$_{8/9}$Cl$_{8}^{+}$,
we used $r_\mathrm{cut} = 5.29$~\AA{} (10 Bohr), 6 Bessel radial functions, $c = 8$, $l_\mathrm{max} = 3$, $\nu_\mathrm{max} = 3$, $N_\mathrm{embedding} = 2$, no message passing,
1-dimensional hidden variable, $\sigma = 1.5$~\AA{}, and $k_c = 2\pi/3$ ($dl=3$~\AA). 

For Au$_{2}$-MgO(001),
we used $r_\mathrm{cut} = 5.5$~\AA, 6 Bessel radial functions, $c = 12$, $l_\mathrm{max} = 3$, $\nu_\mathrm{max} = 3$, $N_\mathrm{embedding} = 4$, no message passing,
1-dimensional hidden variable, $\sigma = 1$~\AA{}, and $k_c = \pi$ ($dl=2$~\AA). 

\paragraph{Electrolyte/solid interfaces}
The Pt(111)/KF(aq) interface dataset from Ref.~\cite{zhu2024machine} was
computed at the PBE-D3 level of theory,
and it contains 4687 configurations covering bulk KF/water electrolytes,
KF/water electrolyte-vapor interfaces,
and KF/water electrolyte-Pt(111) interfaces.

We used a random train/valid/test split of 3318/369/1000 configurations for training the CACE-SR and CACE-LR models.
The CACE-SR model uses $r_\mathrm{cut} = 5.5$~\AA, 6 Bessel radial functions, $c = 12$, $l_\mathrm{max} = 3$, $\nu_\mathrm{max} = 3$, $N_\mathrm{embedding} = 5$, no message passing.
The LR model uses a one-dimensional hidden variable, $\sigma = 1$~\AA{}, and $k_c = \pi$ ($dl=2$~\AA). 

The TiO$_2$(101)/NaCl+NaOH+HCl(aq) dataset from Ref.~\cite{zhang2024electrical} contains a total of 30103 configurations and spans a comprehensive range of gas phase water, bulk solutions, and TiO$_2$, and interfacial configurations.
The dataset was computed at the SCAN DFT level of theory and was collected through an
active learning approach.

We used a random train/valid/test split of 24393/2710/3000 configurations for training the CACE-SR and CACE-LR models.
The CACE-SR model uses $r_\mathrm{cut} = 5.5$~\AA, 6 Bessel radial functions, $c = 12$, $l_\mathrm{max} = 3$, $\nu_\mathrm{max} = 3$, $N_\mathrm{embedding} = 5$, no message passing.
The LR model uses a one-dimensional hidden variable, $\sigma = 1$~\AA{}, and $k_c = \pi$ ($dl=2$~\AA). 

To perform the MD simulation of the TiO$_2$(101)/NaCl(aq) system, we used the same system setup as Ref~\cite{zhang2024electrical}:
The periodic system, illustrated in Fig.~\ref{fig:tio2}, consisting of a five-layer ($3 \times 9$) anatase (101) slab (540 TiO$_2$ units) in contact
with a 67 \AA{} thick layer of aqueous electrolyte (2376 water molecules and 18 NaCl ion pairs).
We used NVT ensemble at 600~K with the Nose-Hoover thermostat.
The timestep was set to 1~fs, and we modified the hydrogen mass to 10.
The total length was 5~ns.

\paragraph{Interphase of LiCl-GaF$_3$}  
To generate the training dataset, we used Bayesian force fields implemented in the \texttt{Flare} package \cite{Vandermause2022_active} to sample the atomic configurations with on-the-fly (OTF) MD simulations of the interface structures of LiCl(001)/GaF$_3$(001), which were generated with the \texttt{CoherentInterfaceBuilder} in \texttt{pymatgen} package \cite{Ong2013_pmg}. The DFT calculation was called when the uncertainty threshold is higher than \texttt{std\_tolerance\_factor=-0.04} in \texttt{Flare}. OTF-MD in the NVT ensemble was initiated from each strained configuration by heating from 0 K to the target temperatures ($T=600/1200$ K). The DFT calculations were performed with VASP in the generalized gradient approximation (GGA) with PBE functional~\cite{perdew1996_GGA}, using a $k$-point mesh of 1000 per reciprocal atom and a plane-wave energy cutoff of 520 eV. The calculations were converged to $10^{-6}$ eV in total energy and the DFT-D3 method of Grimme was used to include Van der Waals corrections \cite{grimme2010_D3}. In total, 3339 DFT-calculated atomic configurations were collected and split into training/validation/test sets with a ratio of 8:1:1.

For the CACE representation, we used 6 Bessel radial functions with $c = 8$, $l_\mathrm{max} = 3$, $\nu_\mathrm{max} = 3$, $N_\mathrm{embedding} = 3$, one message passing, and a cutoff of $r_\mathrm{cut} = 5.5$~\AA. For the long-range component, we used a one-dimensional $q$, $\sigma = 1$~\AA{}, and a maximum cutoff of $k_c = \pi$ ($dl=2$~\AA) in the Ewald summation.

The atomic-resolved force uncertainty was calculated as the root sum of variances along the Cartesian coordinates: $\sigma(F) = \sqrt{\sigma^2(F_x) + \sigma^2(F_y) + \sigma^2(F_z)}$. For each directional component, the variance $\sigma^2(F_i)$ was computed across the ensemble of $N=4$ models using $\sigma^2(F_i) = \frac{1}{N}\sum_{j=1}^N(F_i^j - \Bar{F}_i)^2$, where $F_i^j$ represents the force prediction from the $j$-th model in direction $i \in \{x,y,z\}$, and $\Bar{F}_i$ denotes the ensemble-averaged force in that direction.

\paragraph{Implentation}
We implemented the LES method using PyTorch, and the code is available in 
    \url{https://github.com/BingqingCheng/cace}.
    The raw predicted hidden variables should be scaled by a factor of $1/9.48933$ to obtain the LES charges for e.g. dipole moment prediction, due to the internal normalization factor used ($1/2\epsilon_0=1$).
    
\section{Appendix}

\subsection{Long-range interactions for asymptotic decaying potentials}
\label{appdix:lr_decay}
One choice for the convergence function is~\cite{williams1971accelerated}
\begin{equation}
\varphi(r) = 
\frac{\Gamma(\frac{p}{2}, r^2/2\sigma^2)}{\Gamma(p/2)} 
= \frac{1}{\Gamma(\frac{p}{2})} \int_{r^2/2\sigma^2}^{\infty} t^{\frac{p}{2}-1} e^{-t} dt,
\end{equation}
where $\Gamma(m)$ and $\Gamma(m,x)$ are the gamma function and the incomplete gamma function, respectively. 

The short-range part of $E_p$ can easily be evaluated in the real space, and the long-range part can be computed in the reciprocal space.
The three-dimensional Fourier transform of
\begin{equation}
    \text{FT}_3 \left [\int_V d\mathbf{r}  
    \sum_{i=1}^{N}  q_i \delta(\mathbf{r}_i - \mathbf{r})
    \int_V d\mathbf{r}' \sum_{j=1}^N q_j \delta(\mathbf{r}_j - \mathbf{r}')
    \right] = | S(\mathbf{k}) |^2
\end{equation}
and
\[
\text{FT}_3\left[\frac{1 - \varphi(r)}{r^p}\right] = 
2^{3-p}\pi^{3/2} k^{p-3} \Gamma \left( -\frac{p}{2} + \frac{3}{2}, \frac{ \sigma^2 k^2}{2} \right) 
/ \Gamma \left( \frac{p}{2} \right).
\]
Using Parseval's theorem,
\begin{equation}
    E_p^{lr} =
    \dfrac{2^{2-p}}{V}
    \dfrac{\pi^{3/2}}{ \Gamma \left( \frac{p}{2} \right)}\sum_{k>0}
     k^{p-3} \Gamma \left( -\frac{p}{2} + \frac{3}{2},\frac{ \sigma^2 k^2}{2}  \right) 
    | S(\mathbf{k}) |^2
\end{equation}

\noindent where a factor of 2 is absorbed from double counting pairwise interactions. For $p=1$, as $\Gamma(1/2)=\sqrt{\pi}$ and $\Gamma(1,x) = \exp(-x)$,
\begin{equation}
        E_1^{lr} 
    = \dfrac{2\pi}{V} \sum_{k>0}
    \dfrac{1}{k^2}
    e^{-\sigma^2 k^2/2} |S(\mathbf{k})|^2.
\end{equation}
For $p=6$, 
\begin{multline}
       E_6^{lr} =
\left( \frac{\pi^{3/2}}{24 V} \right) \sum_{k>0} 
k^3 \\
\left[ \pi^{1/2} \mathrm{erfc} (b) + \left( \frac{1}{2b^3} - \frac{1}{b} \right) \times \exp(-b^2) \right] |S(\mathbf{k})|^2 ,
\end{multline}
where $b^2 = \sigma^2 k^2/2$, and $\mathrm{erfc}$ denotes the complimentary error function.

\textbf{Data availability}

The training sets, training scripts, and trained CACE potentials are available at \url{https://github.com/BingqingCheng/cace-lr-fit}.

\textbf{Code availability}
The CACE package is publicly available at \url{https://github.com/BingqingCheng/cace}.

\textbf{Acknowledgements}

We thank Chunyi Zhang for providing the TiO$_2$(101)/NaCl+NaOH+HCl(aq) dataset and for useful discussions.
We thank Jia-Xin Zhu for providing the Pt(111)/KF(aq) dataset.
We thank Tsz Wai Ko and Jonas Finkler for useful discussions and for the DFT-optimized Au$_2$-MgO(001) structures.
We thank Junmin Chen for discussions.
D.K and B.C. acknowledge funding from Toyota Research Institute Synthesis Advanced Research Challenge. D.S.K. and P.Z. acknowledge funding from BIDMaP Postdoctoral Fellowship.

\textbf{Competing Interests}
B.C. has an equity stake in AIMATX Inc.
University of California, Berkeley has filed a provisional patent for the Latent Ewald Summation algorithm.


\begin{thebibliography}{63}%
\makeatletter
\providecommand \@ifxundefined [1]{%
 \@ifx{#1\undefined}
}%
\providecommand \@ifnum [1]{%
 \ifnum #1\expandafter \@firstoftwo
 \else \expandafter \@secondoftwo
 \fi
}%
\providecommand \@ifx [1]{%
 \ifx #1\expandafter \@firstoftwo
 \else \expandafter \@secondoftwo
 \fi
}%
\providecommand \natexlab [1]{#1}%
\providecommand \enquote  [1]{``#1''}%
\providecommand \bibnamefont  [1]{#1}%
\providecommand \bibfnamefont [1]{#1}%
\providecommand \citenamefont [1]{#1}%
\providecommand \href@noop [0]{\@secondoftwo}%
\providecommand \href [0]{\begingroup \@sanitize@url \@href}%
\providecommand \@href[1]{\@@startlink{#1}\@@href}%
\providecommand \@@href[1]{\endgroup#1\@@endlink}%
\providecommand \@sanitize@url [0]{\catcode `\\12\catcode `\$12\catcode `\&12\catcode `\#12\catcode `\^12\catcode `\_12\catcode `\%12\relax}%
\providecommand \@@startlink[1]{}%
\providecommand \@@endlink[0]{}%
\providecommand \url  [0]{\begingroup\@sanitize@url \@url }%
\providecommand \@url [1]{\endgroup\@href {#1}{\urlprefix }}%
\providecommand \urlprefix  [0]{URL }%
\providecommand \Eprint [0]{\href }%
\providecommand \doibase [0]{http://dx.doi.org/}%
\providecommand \selectlanguage [0]{\@gobble}%
\providecommand \bibinfo  [0]{\@secondoftwo}%
\providecommand \bibfield  [0]{\@secondoftwo}%
\providecommand \translation [1]{[#1]}%
\providecommand \BibitemOpen [0]{}%
\providecommand \bibitemStop [0]{}%
\providecommand \bibitemNoStop [0]{.\EOS\space}%
\providecommand \EOS [0]{\spacefactor3000\relax}%
\providecommand \BibitemShut  [1]{\csname bibitem#1\endcsname}%
\let\auto@bib@innerbib\@empty
%</preamble>
\bibitem [{\citenamefont {French}\ \emph {et~al.}(2010)\citenamefont {French}, \citenamefont {Parsegian}, \citenamefont {Podgornik}, \citenamefont {Rajter}, \citenamefont {Jagota}, \citenamefont {Luo}, \citenamefont {Asthagiri}, \citenamefont {Chaudhury}, \citenamefont {Chiang}, \citenamefont {Granick} \emph {et~al.}}]{french2010long}%
  \BibitemOpen
  \bibfield  {author} {\bibinfo {author} {\bibfnamefont {Roger~H}\ \bibnamefont {French}}, \bibinfo {author} {\bibfnamefont {V~Adrian}\ \bibnamefont {Parsegian}}, \bibinfo {author} {\bibfnamefont {Rudolf}\ \bibnamefont {Podgornik}}, \bibinfo {author} {\bibfnamefont {Rick~F}\ \bibnamefont {Rajter}}, \bibinfo {author} {\bibfnamefont {Anand}\ \bibnamefont {Jagota}}, \bibinfo {author} {\bibfnamefont {Jian}\ \bibnamefont {Luo}}, \bibinfo {author} {\bibfnamefont {Dilip}\ \bibnamefont {Asthagiri}}, \bibinfo {author} {\bibfnamefont {Manoj~K}\ \bibnamefont {Chaudhury}}, \bibinfo {author} {\bibfnamefont {Yet-ming}\ \bibnamefont {Chiang}}, \bibinfo {author} {\bibfnamefont {Steve}\ \bibnamefont {Granick}},  \emph {et~al.},\ }\bibfield  {title} {\enquote {\bibinfo {title} {Long range interactions in nanoscale science},}\ }\href@noop {} {\bibfield  {journal} {\bibinfo  {journal} {Reviews of Modern Physics}\ }\textbf {\bibinfo {volume} {82}},\ \bibinfo {pages} {1887--1944} (\bibinfo {year} {2010})}\BibitemShut {NoStop}%
\bibitem [{\citenamefont {Laks}\ \emph {et~al.}(1992)\citenamefont {Laks}, \citenamefont {Ferreira}, \citenamefont {Froyen},\ and\ \citenamefont {Zunger}}]{efficient_CE}%
  \BibitemOpen
  \bibfield  {author} {\bibinfo {author} {\bibfnamefont {David~B.}\ \bibnamefont {Laks}}, \bibinfo {author} {\bibfnamefont {L.~G.}\ \bibnamefont {Ferreira}}, \bibinfo {author} {\bibfnamefont {Sverre}\ \bibnamefont {Froyen}}, \ and\ \bibinfo {author} {\bibfnamefont {Alex}\ \bibnamefont {Zunger}},\ }\bibfield  {title} {\enquote {\bibinfo {title} {Efficient cluster expansion for substitutional systems},}\ }\href@noop {} {\bibfield  {journal} {\bibinfo  {journal} {Phys. Rev. B}\ }\textbf {\bibinfo {volume} {46}},\ \bibinfo {pages} {12587--12605} (\bibinfo {year} {1992})}\BibitemShut {NoStop}%
\bibitem [{\citenamefont {Kirby}\ and\ \citenamefont {Jungwirth}(2019)}]{kirby2019charge}%
  \BibitemOpen
  \bibfield  {author} {\bibinfo {author} {\bibfnamefont {Brian~J}\ \bibnamefont {Kirby}}\ and\ \bibinfo {author} {\bibfnamefont {Pavel}\ \bibnamefont {Jungwirth}},\ }\bibfield  {title} {\enquote {\bibinfo {title} {Charge scaling manifesto: A way of reconciling the inherently macroscopic and microscopic natures of molecular simulations},}\ }\href@noop {} {\bibfield  {journal} {\bibinfo  {journal} {The journal of physical chemistry letters}\ }\textbf {\bibinfo {volume} {10}},\ \bibinfo {pages} {7531--7536} (\bibinfo {year} {2019})}\BibitemShut {NoStop}%
\bibitem [{\citenamefont {Siepmann}\ and\ \citenamefont {Sprik}(1995)}]{siepmann1995influence}%
  \BibitemOpen
  \bibfield  {author} {\bibinfo {author} {\bibfnamefont {J~Ilja}\ \bibnamefont {Siepmann}}\ and\ \bibinfo {author} {\bibfnamefont {Michiel}\ \bibnamefont {Sprik}},\ }\bibfield  {title} {\enquote {\bibinfo {title} {Influence of surface topology and electrostatic potential on water/electrode systems},}\ }\href@noop {} {\bibfield  {journal} {\bibinfo  {journal} {The Journal of chemical physics}\ }\textbf {\bibinfo {volume} {102}},\ \bibinfo {pages} {511--524} (\bibinfo {year} {1995})}\BibitemShut {NoStop}%
\bibitem [{\citenamefont {Keith}\ \emph {et~al.}(2021)\citenamefont {Keith}, \citenamefont {Vassilev-Galindo}, \citenamefont {Cheng}, \citenamefont {Chmiela}, \citenamefont {Gastegger}, \citenamefont {M{\"u}ller},\ and\ \citenamefont {Tkatchenko}}]{keith2021combining}%
  \BibitemOpen
  \bibfield  {author} {\bibinfo {author} {\bibfnamefont {John~A}\ \bibnamefont {Keith}}, \bibinfo {author} {\bibfnamefont {Valentin}\ \bibnamefont {Vassilev-Galindo}}, \bibinfo {author} {\bibfnamefont {Bingqing}\ \bibnamefont {Cheng}}, \bibinfo {author} {\bibfnamefont {Stefan}\ \bibnamefont {Chmiela}}, \bibinfo {author} {\bibfnamefont {Michael}\ \bibnamefont {Gastegger}}, \bibinfo {author} {\bibfnamefont {Klaus-Robert}\ \bibnamefont {M{\"u}ller}}, \ and\ \bibinfo {author} {\bibfnamefont {Alexandre}\ \bibnamefont {Tkatchenko}},\ }\bibfield  {title} {\enquote {\bibinfo {title} {Combining machine learning and computational chemistry for predictive insights into chemical systems},}\ }\href@noop {} {\bibfield  {journal} {\bibinfo  {journal} {Chemical reviews}\ }\textbf {\bibinfo {volume} {121}},\ \bibinfo {pages} {9816--9872} (\bibinfo {year} {2021})}\BibitemShut {NoStop}%
\bibitem [{\citenamefont {Unke}\ \emph {et~al.}(2021)\citenamefont {Unke}, \citenamefont {Chmiela}, \citenamefont {Sauceda}, \citenamefont {Gastegger}, \citenamefont {Poltavsky}, \citenamefont {Sch{\"u}tt}, \citenamefont {Tkatchenko},\ and\ \citenamefont {M{\"u}ller}}]{unke2021machine}%
  \BibitemOpen
  \bibfield  {author} {\bibinfo {author} {\bibfnamefont {Oliver~T}\ \bibnamefont {Unke}}, \bibinfo {author} {\bibfnamefont {Stefan}\ \bibnamefont {Chmiela}}, \bibinfo {author} {\bibfnamefont {Huziel~E}\ \bibnamefont {Sauceda}}, \bibinfo {author} {\bibfnamefont {Michael}\ \bibnamefont {Gastegger}}, \bibinfo {author} {\bibfnamefont {Igor}\ \bibnamefont {Poltavsky}}, \bibinfo {author} {\bibfnamefont {Kristof~T}\ \bibnamefont {Sch{\"u}tt}}, \bibinfo {author} {\bibfnamefont {Alexandre}\ \bibnamefont {Tkatchenko}}, \ and\ \bibinfo {author} {\bibfnamefont {Klaus-Robert}\ \bibnamefont {M{\"u}ller}},\ }\bibfield  {title} {\enquote {\bibinfo {title} {Machine learning force fields},}\ }\href@noop {} {\bibfield  {journal} {\bibinfo  {journal} {Chemical Reviews}\ }\textbf {\bibinfo {volume} {121}},\ \bibinfo {pages} {10142--10186} (\bibinfo {year} {2021})}\BibitemShut {NoStop}%
\bibitem [{\citenamefont {Niblett}\ \emph {et~al.}(2021)\citenamefont {Niblett}, \citenamefont {Galib},\ and\ \citenamefont {Limmer}}]{niblett2021learning}%
  \BibitemOpen
  \bibfield  {author} {\bibinfo {author} {\bibfnamefont {Samuel~P}\ \bibnamefont {Niblett}}, \bibinfo {author} {\bibfnamefont {Mirza}\ \bibnamefont {Galib}}, \ and\ \bibinfo {author} {\bibfnamefont {David~T}\ \bibnamefont {Limmer}},\ }\bibfield  {title} {\enquote {\bibinfo {title} {Learning intermolecular forces at liquid--vapor interfaces},}\ }\href@noop {} {\bibfield  {journal} {\bibinfo  {journal} {The Journal of chemical physics}\ }\textbf {\bibinfo {volume} {155}} (\bibinfo {year} {2021})}\BibitemShut {NoStop}%
\bibitem [{\citenamefont {Grisafi}\ and\ \citenamefont {Ceriotti}(2019)}]{grisafi2019incorporating}%
  \BibitemOpen
  \bibfield  {author} {\bibinfo {author} {\bibfnamefont {Andrea}\ \bibnamefont {Grisafi}}\ and\ \bibinfo {author} {\bibfnamefont {Michele}\ \bibnamefont {Ceriotti}},\ }\bibfield  {title} {\enquote {\bibinfo {title} {Incorporating long-range physics in atomic-scale machine learning},}\ }\href@noop {} {\bibfield  {journal} {\bibinfo  {journal} {The Journal of chemical physics}\ }\textbf {\bibinfo {volume} {151}} (\bibinfo {year} {2019})}\BibitemShut {NoStop}%
\bibitem [{\citenamefont {Huguenin-Dumittan}\ \emph {et~al.}(2023)\citenamefont {Huguenin-Dumittan}, \citenamefont {Loche}, \citenamefont {Haoran},\ and\ \citenamefont {Ceriotti}}]{huguenin2023physics}%
  \BibitemOpen
  \bibfield  {author} {\bibinfo {author} {\bibfnamefont {Kevin~K}\ \bibnamefont {Huguenin-Dumittan}}, \bibinfo {author} {\bibfnamefont {Philip}\ \bibnamefont {Loche}}, \bibinfo {author} {\bibfnamefont {Ni}~\bibnamefont {Haoran}}, \ and\ \bibinfo {author} {\bibfnamefont {Michele}\ \bibnamefont {Ceriotti}},\ }\bibfield  {title} {\enquote {\bibinfo {title} {Physics-inspired equivariant descriptors of nonbonded interactions},}\ }\href@noop {} {\bibfield  {journal} {\bibinfo  {journal} {The Journal of Physical Chemistry Letters}\ }\textbf {\bibinfo {volume} {14}},\ \bibinfo {pages} {9612--9618} (\bibinfo {year} {2023})}\BibitemShut {NoStop}%
\bibitem [{\citenamefont {Zhang}\ \emph {et~al.}(2022)\citenamefont {Zhang}, \citenamefont {Wang}, \citenamefont {Muniz}, \citenamefont {Panagiotopoulos}, \citenamefont {Car} \emph {et~al.}}]{zhang2022deep}%
  \BibitemOpen
  \bibfield  {author} {\bibinfo {author} {\bibfnamefont {Linfeng}\ \bibnamefont {Zhang}}, \bibinfo {author} {\bibfnamefont {Han}\ \bibnamefont {Wang}}, \bibinfo {author} {\bibfnamefont {Maria~Carolina}\ \bibnamefont {Muniz}}, \bibinfo {author} {\bibfnamefont {Athanassios~Z}\ \bibnamefont {Panagiotopoulos}}, \bibinfo {author} {\bibfnamefont {Roberto}\ \bibnamefont {Car}},  \emph {et~al.},\ }\bibfield  {title} {\enquote {\bibinfo {title} {A deep potential model with long-range electrostatic interactions},}\ }\href@noop {} {\bibfield  {journal} {\bibinfo  {journal} {The Journal of Chemical Physics}\ }\textbf {\bibinfo {volume} {156}} (\bibinfo {year} {2022})}\BibitemShut {NoStop}%
\bibitem [{\citenamefont {Monacelli}\ and\ \citenamefont {Marzari}(2024)}]{monacelli2024electrostatic}%
  \BibitemOpen
  \bibfield  {author} {\bibinfo {author} {\bibfnamefont {Lorenzo}\ \bibnamefont {Monacelli}}\ and\ \bibinfo {author} {\bibfnamefont {Nicola}\ \bibnamefont {Marzari}},\ }\bibfield  {title} {\enquote {\bibinfo {title} {Electrostatic interactions in atomistic and machine-learned potentials for polar materials},}\ }\href@noop {} {\bibfield  {journal} {\bibinfo  {journal} {arXiv preprint arXiv:2412.01642}\ } (\bibinfo {year} {2024})}\BibitemShut {NoStop}%
\bibitem [{\citenamefont {Ko}\ \emph {et~al.}(2021)\citenamefont {Ko}, \citenamefont {Finkler}, \citenamefont {Goedecker},\ and\ \citenamefont {Behler}}]{ko2021fourth}%
  \BibitemOpen
  \bibfield  {author} {\bibinfo {author} {\bibfnamefont {Tsz~Wai}\ \bibnamefont {Ko}}, \bibinfo {author} {\bibfnamefont {Jonas~A}\ \bibnamefont {Finkler}}, \bibinfo {author} {\bibfnamefont {Stefan}\ \bibnamefont {Goedecker}}, \ and\ \bibinfo {author} {\bibfnamefont {J{\"o}rg}\ \bibnamefont {Behler}},\ }\bibfield  {title} {\enquote {\bibinfo {title} {A fourth-generation high-dimensional neural network potential with accurate electrostatics including non-local charge transfer},}\ }\href@noop {} {\bibfield  {journal} {\bibinfo  {journal} {Nature communications}\ }\textbf {\bibinfo {volume} {12}},\ \bibinfo {pages} {398} (\bibinfo {year} {2021})}\BibitemShut {NoStop}%
\bibitem [{\citenamefont {Unke}\ and\ \citenamefont {Meuwly}(2019)}]{unke2019physnet}%
  \BibitemOpen
  \bibfield  {author} {\bibinfo {author} {\bibfnamefont {Oliver~T}\ \bibnamefont {Unke}}\ and\ \bibinfo {author} {\bibfnamefont {Markus}\ \bibnamefont {Meuwly}},\ }\bibfield  {title} {\enquote {\bibinfo {title} {Physnet: A neural network for predicting energies, forces, dipole moments, and partial charges},}\ }\href@noop {} {\bibfield  {journal} {\bibinfo  {journal} {Journal of chemical theory and computation}\ }\textbf {\bibinfo {volume} {15}},\ \bibinfo {pages} {3678--3693} (\bibinfo {year} {2019})}\BibitemShut {NoStop}%
\bibitem [{\citenamefont {Gao}\ and\ \citenamefont {Remsing}(2022)}]{gao2022self}%
  \BibitemOpen
  \bibfield  {author} {\bibinfo {author} {\bibfnamefont {Ang}\ \bibnamefont {Gao}}\ and\ \bibinfo {author} {\bibfnamefont {Richard~C}\ \bibnamefont {Remsing}},\ }\bibfield  {title} {\enquote {\bibinfo {title} {Self-consistent determination of long-range electrostatics in neural network potentials},}\ }\href@noop {} {\bibfield  {journal} {\bibinfo  {journal} {Nature communications}\ }\textbf {\bibinfo {volume} {13}},\ \bibinfo {pages} {1572} (\bibinfo {year} {2022})}\BibitemShut {NoStop}%
\bibitem [{\citenamefont {Sifain}\ \emph {et~al.}(2018)\citenamefont {Sifain}, \citenamefont {Lubbers}, \citenamefont {Nebgen}, \citenamefont {Smith}, \citenamefont {Lokhov}, \citenamefont {Isayev}, \citenamefont {Roitberg}, \citenamefont {Barros},\ and\ \citenamefont {Tretiak}}]{sifain2018discovering}%
  \BibitemOpen
  \bibfield  {author} {\bibinfo {author} {\bibfnamefont {Andrew~E}\ \bibnamefont {Sifain}}, \bibinfo {author} {\bibfnamefont {Nicholas}\ \bibnamefont {Lubbers}}, \bibinfo {author} {\bibfnamefont {Benjamin~T}\ \bibnamefont {Nebgen}}, \bibinfo {author} {\bibfnamefont {Justin~S}\ \bibnamefont {Smith}}, \bibinfo {author} {\bibfnamefont {Andrey~Y}\ \bibnamefont {Lokhov}}, \bibinfo {author} {\bibfnamefont {Olexandr}\ \bibnamefont {Isayev}}, \bibinfo {author} {\bibfnamefont {Adrian~E}\ \bibnamefont {Roitberg}}, \bibinfo {author} {\bibfnamefont {Kipton}\ \bibnamefont {Barros}}, \ and\ \bibinfo {author} {\bibfnamefont {Sergei}\ \bibnamefont {Tretiak}},\ }\bibfield  {title} {\enquote {\bibinfo {title} {Discovering a transferable charge assignment model using machine learning},}\ }\href@noop {} {\bibfield  {journal} {\bibinfo  {journal} {The journal of physical chemistry letters}\ }\textbf {\bibinfo {volume} {9}},\ \bibinfo {pages} {4495--4501} (\bibinfo {year} {2018})}\BibitemShut {NoStop}%
\bibitem [{\citenamefont {Gong}\ \emph {et~al.}(2024)\citenamefont {Gong}, \citenamefont {Zhang}, \citenamefont {Mu}, \citenamefont {Pu}, \citenamefont {Wang}, \citenamefont {Yu}, \citenamefont {Chen}, \citenamefont {Zheng}, \citenamefont {Wang}, \citenamefont {Chen} \emph {et~al.}}]{gong2024bamboo}%
  \BibitemOpen
  \bibfield  {author} {\bibinfo {author} {\bibfnamefont {Sheng}\ \bibnamefont {Gong}}, \bibinfo {author} {\bibfnamefont {Yumin}\ \bibnamefont {Zhang}}, \bibinfo {author} {\bibfnamefont {Zhenliang}\ \bibnamefont {Mu}}, \bibinfo {author} {\bibfnamefont {Zhichen}\ \bibnamefont {Pu}}, \bibinfo {author} {\bibfnamefont {Hongyi}\ \bibnamefont {Wang}}, \bibinfo {author} {\bibfnamefont {Zhiao}\ \bibnamefont {Yu}}, \bibinfo {author} {\bibfnamefont {Mengyi}\ \bibnamefont {Chen}}, \bibinfo {author} {\bibfnamefont {Tianze}\ \bibnamefont {Zheng}}, \bibinfo {author} {\bibfnamefont {Zhi}\ \bibnamefont {Wang}}, \bibinfo {author} {\bibfnamefont {Lifei}\ \bibnamefont {Chen}},  \emph {et~al.},\ }\bibfield  {title} {\enquote {\bibinfo {title} {Bamboo: a predictive and transferable machine learning force field framework for liquid electrolyte development},}\ }\href@noop {} {\bibfield  {journal} {\bibinfo  {journal} {arXiv preprint arXiv:2404.07181}\ } (\bibinfo {year} {2024})}\BibitemShut {NoStop}%
\bibitem [{\citenamefont {Shaidu}\ \emph {et~al.}(2024)\citenamefont {Shaidu}, \citenamefont {Pellegrini}, \citenamefont {K{\"u}{\c{c}}{\"u}kbenli}, \citenamefont {Lot},\ and\ \citenamefont {de~Gironcoli}}]{shaidu2024incorporating}%
  \BibitemOpen
  \bibfield  {author} {\bibinfo {author} {\bibfnamefont {Yusuf}\ \bibnamefont {Shaidu}}, \bibinfo {author} {\bibfnamefont {Franco}\ \bibnamefont {Pellegrini}}, \bibinfo {author} {\bibfnamefont {Emine}\ \bibnamefont {K{\"u}{\c{c}}{\"u}kbenli}}, \bibinfo {author} {\bibfnamefont {Ruggero}\ \bibnamefont {Lot}}, \ and\ \bibinfo {author} {\bibfnamefont {Stefano}\ \bibnamefont {de~Gironcoli}},\ }\bibfield  {title} {\enquote {\bibinfo {title} {Incorporating long-range electrostatics in neural network potentials via variational charge equilibration from shortsighted ingredients},}\ }\href@noop {} {\bibfield  {journal} {\bibinfo  {journal} {npj Computational Materials}\ }\textbf {\bibinfo {volume} {10}},\ \bibinfo {pages} {47} (\bibinfo {year} {2024})}\BibitemShut {NoStop}%
\bibitem [{\citenamefont {Rappe}\ and\ \citenamefont {Goddard~III}(1991)}]{rappe1991charge}%
  \BibitemOpen
  \bibfield  {author} {\bibinfo {author} {\bibfnamefont {Anthony~K}\ \bibnamefont {Rappe}}\ and\ \bibinfo {author} {\bibfnamefont {William~A}\ \bibnamefont {Goddard~III}},\ }\bibfield  {title} {\enquote {\bibinfo {title} {Charge equilibration for molecular dynamics simulations},}\ }\href@noop {} {\bibfield  {journal} {\bibinfo  {journal} {The Journal of Physical Chemistry}\ }\textbf {\bibinfo {volume} {95}},\ \bibinfo {pages} {3358--3363} (\bibinfo {year} {1991})}\BibitemShut {NoStop}%
\bibitem [{\citenamefont {Yu}\ \emph {et~al.}(2022)\citenamefont {Yu}, \citenamefont {Hong}, \citenamefont {Chen}, \citenamefont {Gong},\ and\ \citenamefont {Xiang}}]{yu2022capturing}%
  \BibitemOpen
  \bibfield  {author} {\bibinfo {author} {\bibfnamefont {Hongyu}\ \bibnamefont {Yu}}, \bibinfo {author} {\bibfnamefont {Liangliang}\ \bibnamefont {Hong}}, \bibinfo {author} {\bibfnamefont {Shiyou}\ \bibnamefont {Chen}}, \bibinfo {author} {\bibfnamefont {Xingao}\ \bibnamefont {Gong}}, \ and\ \bibinfo {author} {\bibfnamefont {Hongjun}\ \bibnamefont {Xiang}},\ }\bibfield  {title} {\enquote {\bibinfo {title} {Capturing long-range interaction with reciprocal space neural network},}\ }\href@noop {} {\bibfield  {journal} {\bibinfo  {journal} {arXiv preprint arXiv:2211.16684}\ } (\bibinfo {year} {2022})}\BibitemShut {NoStop}%
\bibitem [{\citenamefont {Kosmala}\ \emph {et~al.}(2023)\citenamefont {Kosmala}, \citenamefont {Gasteiger}, \citenamefont {Gao},\ and\ \citenamefont {G{\"u}nnemann}}]{kosmala2023ewald}%
  \BibitemOpen
  \bibfield  {author} {\bibinfo {author} {\bibfnamefont {Arthur}\ \bibnamefont {Kosmala}}, \bibinfo {author} {\bibfnamefont {Johannes}\ \bibnamefont {Gasteiger}}, \bibinfo {author} {\bibfnamefont {Nicholas}\ \bibnamefont {Gao}}, \ and\ \bibinfo {author} {\bibfnamefont {Stephan}\ \bibnamefont {G{\"u}nnemann}},\ }\bibfield  {title} {\enquote {\bibinfo {title} {Ewald-based long-range message passing for molecular graphs},}\ }in\ \href@noop {} {\emph {\bibinfo {booktitle} {International Conference on Machine Learning}}}\ (\bibinfo {organization} {PMLR},\ \bibinfo {year} {2023})\ pp.\ \bibinfo {pages} {17544--17563}\BibitemShut {NoStop}%
\bibitem [{\citenamefont {Faller}\ \emph {et~al.}(2024)\citenamefont {Faller}, \citenamefont {Kaltak},\ and\ \citenamefont {Kresse}}]{faller2024density}%
  \BibitemOpen
  \bibfield  {author} {\bibinfo {author} {\bibfnamefont {Carolin}\ \bibnamefont {Faller}}, \bibinfo {author} {\bibfnamefont {Merzuk}\ \bibnamefont {Kaltak}}, \ and\ \bibinfo {author} {\bibfnamefont {Georg}\ \bibnamefont {Kresse}},\ }\bibfield  {title} {\enquote {\bibinfo {title} {Density-based long-range electrostatic descriptors for machine learning force fields},}\ }\href@noop {} {\bibfield  {journal} {\bibinfo  {journal} {arXiv preprint arXiv:2406.17595}\ } (\bibinfo {year} {2024})}\BibitemShut {NoStop}%
\bibitem [{\citenamefont {Loche}\ \emph {et~al.}(2024)\citenamefont {Loche}, \citenamefont {Huguenin-Dumittan}, \citenamefont {Honarmand}, \citenamefont {Xu}, \citenamefont {Rumiantsev}, \citenamefont {How}, \citenamefont {Langer},\ and\ \citenamefont {Ceriotti}}]{loche2024fast}%
  \BibitemOpen
  \bibfield  {author} {\bibinfo {author} {\bibfnamefont {Philip}\ \bibnamefont {Loche}}, \bibinfo {author} {\bibfnamefont {Kevin~K}\ \bibnamefont {Huguenin-Dumittan}}, \bibinfo {author} {\bibfnamefont {Melika}\ \bibnamefont {Honarmand}}, \bibinfo {author} {\bibfnamefont {Qianjun}\ \bibnamefont {Xu}}, \bibinfo {author} {\bibfnamefont {Egor}\ \bibnamefont {Rumiantsev}}, \bibinfo {author} {\bibfnamefont {Wei~Bin}\ \bibnamefont {How}}, \bibinfo {author} {\bibfnamefont {Marcel~F}\ \bibnamefont {Langer}}, \ and\ \bibinfo {author} {\bibfnamefont {Michele}\ \bibnamefont {Ceriotti}},\ }\bibfield  {title} {\enquote {\bibinfo {title} {Fast and flexible range-separated models for atomistic machine learning},}\ }\href@noop {} {\bibfield  {journal} {\bibinfo  {journal} {arXiv preprint arXiv:2412.03281}\ } (\bibinfo {year} {2024})}\BibitemShut {NoStop}%
\bibitem [{\citenamefont {Cheng}(2024{\natexlab{a}})}]{cheng2024latent}%
  \BibitemOpen
  \bibfield  {author} {\bibinfo {author} {\bibfnamefont {Bingqing}\ \bibnamefont {Cheng}},\ }\bibfield  {title} {\enquote {\bibinfo {title} {Latent ewald summation for machine learning of long-range interactions},}\ }\href@noop {} {\bibfield  {journal} {\bibinfo  {journal} {arXiv preprint arXiv:2408.15165}\ } (\bibinfo {year} {2024}{\natexlab{a}})}\BibitemShut {NoStop}%
\bibitem [{\citenamefont {Behler}\ and\ \citenamefont {Parrinello}(2007)}]{behler2007generalized}%
  \BibitemOpen
  \bibfield  {author} {\bibinfo {author} {\bibfnamefont {J{\"o}rg}\ \bibnamefont {Behler}}\ and\ \bibinfo {author} {\bibfnamefont {Michele}\ \bibnamefont {Parrinello}},\ }\bibfield  {title} {\enquote {\bibinfo {title} {Generalized neural-network representation of high-dimensional potential-energy surfaces},}\ }\href@noop {} {\bibfield  {journal} {\bibinfo  {journal} {Physical review letters}\ }\textbf {\bibinfo {volume} {98}},\ \bibinfo {pages} {146401} (\bibinfo {year} {2007})}\BibitemShut {NoStop}%
\bibitem [{\citenamefont {Bart{\'o}k}\ \emph {et~al.}(2010)\citenamefont {Bart{\'o}k}, \citenamefont {Payne}, \citenamefont {Kondor},\ and\ \citenamefont {Cs{\'a}nyi}}]{bartok2010gaussian}%
  \BibitemOpen
  \bibfield  {author} {\bibinfo {author} {\bibfnamefont {Albert~P}\ \bibnamefont {Bart{\'o}k}}, \bibinfo {author} {\bibfnamefont {Mike~C}\ \bibnamefont {Payne}}, \bibinfo {author} {\bibfnamefont {Risi}\ \bibnamefont {Kondor}}, \ and\ \bibinfo {author} {\bibfnamefont {G{\'a}bor}\ \bibnamefont {Cs{\'a}nyi}},\ }\bibfield  {title} {\enquote {\bibinfo {title} {Gaussian approximation potentials: The accuracy of quantum mechanics, without the electrons},}\ }\href@noop {} {\bibfield  {journal} {\bibinfo  {journal} {Physical review letters}\ }\textbf {\bibinfo {volume} {104}},\ \bibinfo {pages} {136403} (\bibinfo {year} {2010})}\BibitemShut {NoStop}%
\bibitem [{\citenamefont {Shapeev}(2016)}]{shapeev2016moment}%
  \BibitemOpen
  \bibfield  {author} {\bibinfo {author} {\bibfnamefont {Alexander~V}\ \bibnamefont {Shapeev}},\ }\bibfield  {title} {\enquote {\bibinfo {title} {Moment tensor potentials: A class of systematically improvable interatomic potentials},}\ }\href@noop {} {\bibfield  {journal} {\bibinfo  {journal} {Multiscale Modeling \& Simulation}\ }\textbf {\bibinfo {volume} {14}},\ \bibinfo {pages} {1153--1173} (\bibinfo {year} {2016})}\BibitemShut {NoStop}%
\bibitem [{\citenamefont {Drautz}(2019)}]{drautz2019atomic}%
  \BibitemOpen
  \bibfield  {author} {\bibinfo {author} {\bibfnamefont {Ralf}\ \bibnamefont {Drautz}},\ }\bibfield  {title} {\enquote {\bibinfo {title} {Atomic cluster expansion for accurate and transferable interatomic potentials},}\ }\href@noop {} {\bibfield  {journal} {\bibinfo  {journal} {Physical Review B}\ }\textbf {\bibinfo {volume} {99}},\ \bibinfo {pages} {014104} (\bibinfo {year} {2019})}\BibitemShut {NoStop}%
\bibitem [{\citenamefont {Batzner}\ \emph {et~al.}(2022)\citenamefont {Batzner}, \citenamefont {Musaelian}, \citenamefont {Sun}, \citenamefont {Geiger}, \citenamefont {Mailoa}, \citenamefont {Kornbluth}, \citenamefont {Molinari}, \citenamefont {Smidt},\ and\ \citenamefont {Kozinsky}}]{batzner20223}%
  \BibitemOpen
  \bibfield  {author} {\bibinfo {author} {\bibfnamefont {Simon}\ \bibnamefont {Batzner}}, \bibinfo {author} {\bibfnamefont {Albert}\ \bibnamefont {Musaelian}}, \bibinfo {author} {\bibfnamefont {Lixin}\ \bibnamefont {Sun}}, \bibinfo {author} {\bibfnamefont {Mario}\ \bibnamefont {Geiger}}, \bibinfo {author} {\bibfnamefont {Jonathan~P}\ \bibnamefont {Mailoa}}, \bibinfo {author} {\bibfnamefont {Mordechai}\ \bibnamefont {Kornbluth}}, \bibinfo {author} {\bibfnamefont {Nicola}\ \bibnamefont {Molinari}}, \bibinfo {author} {\bibfnamefont {Tess~E}\ \bibnamefont {Smidt}}, \ and\ \bibinfo {author} {\bibfnamefont {Boris}\ \bibnamefont {Kozinsky}},\ }\bibfield  {title} {\enquote {\bibinfo {title} {E (3)-equivariant graph neural networks for data-efficient and accurate interatomic potentials},}\ }\href@noop {} {\bibfield  {journal} {\bibinfo  {journal} {Nature communications}\ }\textbf {\bibinfo {volume} {13}},\ \bibinfo {pages} {2453} (\bibinfo {year} {2022})}\BibitemShut {NoStop}%
\bibitem [{\citenamefont {Batatia}\ \emph {et~al.}(2022)\citenamefont {Batatia}, \citenamefont {Kovacs}, \citenamefont {Simm}, \citenamefont {Ortner},\ and\ \citenamefont {Cs{\'a}nyi}}]{batatia2022mace}%
  \BibitemOpen
  \bibfield  {author} {\bibinfo {author} {\bibfnamefont {Ilyes}\ \bibnamefont {Batatia}}, \bibinfo {author} {\bibfnamefont {David~P}\ \bibnamefont {Kovacs}}, \bibinfo {author} {\bibfnamefont {Gregor}\ \bibnamefont {Simm}}, \bibinfo {author} {\bibfnamefont {Christoph}\ \bibnamefont {Ortner}}, \ and\ \bibinfo {author} {\bibfnamefont {G{\'a}bor}\ \bibnamefont {Cs{\'a}nyi}},\ }\bibfield  {title} {\enquote {\bibinfo {title} {Mace: Higher order equivariant message passing neural networks for fast and accurate force fields},}\ }\href@noop {} {\bibfield  {journal} {\bibinfo  {journal} {Advances in Neural Information Processing Systems}\ }\textbf {\bibinfo {volume} {35}},\ \bibinfo {pages} {11423--11436} (\bibinfo {year} {2022})}\BibitemShut {NoStop}%
\bibitem [{\citenamefont {Cheng}(2024{\natexlab{b}})}]{cheng2024cartesian}%
  \BibitemOpen
  \bibfield  {author} {\bibinfo {author} {\bibfnamefont {Bingqing}\ \bibnamefont {Cheng}},\ }\bibfield  {title} {\enquote {\bibinfo {title} {Cartesian atomic cluster expansion for machine learning interatomic potentials},}\ }\href@noop {} {\bibfield  {journal} {\bibinfo  {journal} {npj Computational Materials}\ }\textbf {\bibinfo {volume} {10}},\ \bibinfo {pages} {157} (\bibinfo {year} {2024}{\natexlab{b}})}\BibitemShut {NoStop}%
\bibitem [{\citenamefont {Sch{\"u}tt}\ \emph {et~al.}(2017)\citenamefont {Sch{\"u}tt}, \citenamefont {Kindermans}, \citenamefont {Sauceda~Felix}, \citenamefont {Chmiela}, \citenamefont {Tkatchenko},\ and\ \citenamefont {M{\"u}ller}}]{schutt2017schnet}%
  \BibitemOpen
  \bibfield  {author} {\bibinfo {author} {\bibfnamefont {Kristof}\ \bibnamefont {Sch{\"u}tt}}, \bibinfo {author} {\bibfnamefont {Pieter-Jan}\ \bibnamefont {Kindermans}}, \bibinfo {author} {\bibfnamefont {Huziel~Enoc}\ \bibnamefont {Sauceda~Felix}}, \bibinfo {author} {\bibfnamefont {Stefan}\ \bibnamefont {Chmiela}}, \bibinfo {author} {\bibfnamefont {Alexandre}\ \bibnamefont {Tkatchenko}}, \ and\ \bibinfo {author} {\bibfnamefont {Klaus-Robert}\ \bibnamefont {M{\"u}ller}},\ }\bibfield  {title} {\enquote {\bibinfo {title} {Schnet: A continuous-filter convolutional neural network for modeling quantum interactions},}\ }\href@noop {} {\bibfield  {journal} {\bibinfo  {journal} {Advances in neural information processing systems}\ }\textbf {\bibinfo {volume} {30}} (\bibinfo {year} {2017})}\BibitemShut {NoStop}%
\bibitem [{\citenamefont {Deng}\ \emph {et~al.}(2023)\citenamefont {Deng}, \citenamefont {Zhong}, \citenamefont {Jun}, \citenamefont {Riebesell}, \citenamefont {Han}, \citenamefont {Bartel},\ and\ \citenamefont {Ceder}}]{deng2023chgnet}%
  \BibitemOpen
  \bibfield  {author} {\bibinfo {author} {\bibfnamefont {Bowen}\ \bibnamefont {Deng}}, \bibinfo {author} {\bibfnamefont {Peichen}\ \bibnamefont {Zhong}}, \bibinfo {author} {\bibfnamefont {KyuJung}\ \bibnamefont {Jun}}, \bibinfo {author} {\bibfnamefont {Janosh}\ \bibnamefont {Riebesell}}, \bibinfo {author} {\bibfnamefont {Kevin}\ \bibnamefont {Han}}, \bibinfo {author} {\bibfnamefont {Christopher~J}\ \bibnamefont {Bartel}}, \ and\ \bibinfo {author} {\bibfnamefont {Gerbrand}\ \bibnamefont {Ceder}},\ }\bibfield  {title} {\enquote {\bibinfo {title} {Chgnet as a pretrained universal neural network potential for charge-informed atomistic modelling},}\ }\href@noop {} {\bibfield  {journal} {\bibinfo  {journal} {Nature Machine Intelligence}\ }\textbf {\bibinfo {volume} {5}},\ \bibinfo {pages} {1031--1041} (\bibinfo {year} {2023})}\BibitemShut {NoStop}%
\bibitem [{\citenamefont {Haghighatlari}\ \emph {et~al.}(2022)\citenamefont {Haghighatlari}, \citenamefont {Li}, \citenamefont {Guan}, \citenamefont {Zhang}, \citenamefont {Das}, \citenamefont {Stein}, \citenamefont {Heidar-Zadeh}, \citenamefont {Liu}, \citenamefont {Head-Gordon}, \citenamefont {Bertels} \emph {et~al.}}]{haghighatlari2022newtonnet}%
  \BibitemOpen
  \bibfield  {author} {\bibinfo {author} {\bibfnamefont {Mojtaba}\ \bibnamefont {Haghighatlari}}, \bibinfo {author} {\bibfnamefont {Jie}\ \bibnamefont {Li}}, \bibinfo {author} {\bibfnamefont {Xingyi}\ \bibnamefont {Guan}}, \bibinfo {author} {\bibfnamefont {Oufan}\ \bibnamefont {Zhang}}, \bibinfo {author} {\bibfnamefont {Akshaya}\ \bibnamefont {Das}}, \bibinfo {author} {\bibfnamefont {Christopher~J}\ \bibnamefont {Stein}}, \bibinfo {author} {\bibfnamefont {Farnaz}\ \bibnamefont {Heidar-Zadeh}}, \bibinfo {author} {\bibfnamefont {Meili}\ \bibnamefont {Liu}}, \bibinfo {author} {\bibfnamefont {Martin}\ \bibnamefont {Head-Gordon}}, \bibinfo {author} {\bibfnamefont {Luke}\ \bibnamefont {Bertels}},  \emph {et~al.},\ }\bibfield  {title} {\enquote {\bibinfo {title} {Newtonnet: A newtonian message passing network for deep learning of interatomic potentials and forces},}\ }\href@noop {} {\bibfield  {journal} {\bibinfo  {journal} {Digital Discovery}\ }\textbf {\bibinfo {volume} {1}},\ \bibinfo {pages} {333--343} (\bibinfo
  {year} {2022})}\BibitemShut {NoStop}%
\bibitem [{\citenamefont {Williams}(1971)}]{williams1971accelerated}%
  \BibitemOpen
  \bibfield  {author} {\bibinfo {author} {\bibfnamefont {Donald~E}\ \bibnamefont {Williams}},\ }\bibfield  {title} {\enquote {\bibinfo {title} {Accelerated convergence of crystal-lattice potential sums},}\ }\href@noop {} {\bibfield  {journal} {\bibinfo  {journal} {Acta Crystallographica Section A: Crystal Physics, Diffraction, Theoretical and General Crystallography}\ }\textbf {\bibinfo {volume} {27}},\ \bibinfo {pages} {452--455} (\bibinfo {year} {1971})}\BibitemShut {NoStop}%
\bibitem [{\citenamefont {Prodan}\ and\ \citenamefont {Kohn}(2005)}]{prodan2005nearsightedness}%
  \BibitemOpen
  \bibfield  {author} {\bibinfo {author} {\bibfnamefont {Emil}\ \bibnamefont {Prodan}}\ and\ \bibinfo {author} {\bibfnamefont {Walter}\ \bibnamefont {Kohn}},\ }\bibfield  {title} {\enquote {\bibinfo {title} {Nearsightedness of electronic matter},}\ }\href@noop {} {\bibfield  {journal} {\bibinfo  {journal} {Proceedings of the National Academy of Sciences}\ }\textbf {\bibinfo {volume} {102}},\ \bibinfo {pages} {11635--11638} (\bibinfo {year} {2005})}\BibitemShut {NoStop}%
\bibitem [{\citenamefont {Wu}\ \emph {et~al.}(2006)\citenamefont {Wu}, \citenamefont {Tepper},\ and\ \citenamefont {Voth}}]{wu2006flexible}%
  \BibitemOpen
  \bibfield  {author} {\bibinfo {author} {\bibfnamefont {Yujie}\ \bibnamefont {Wu}}, \bibinfo {author} {\bibfnamefont {Harald~L}\ \bibnamefont {Tepper}}, \ and\ \bibinfo {author} {\bibfnamefont {Gregory~A}\ \bibnamefont {Voth}},\ }\bibfield  {title} {\enquote {\bibinfo {title} {Flexible simple point-charge water model with improved liquid-state properties},}\ }\href@noop {} {\bibfield  {journal} {\bibinfo  {journal} {The Journal of chemical physics}\ }\textbf {\bibinfo {volume} {124}} (\bibinfo {year} {2006})}\BibitemShut {NoStop}%
\bibitem [{\citenamefont {Joung}\ and\ \citenamefont {Cheatham~III}(2008)}]{joung2008determination}%
  \BibitemOpen
  \bibfield  {author} {\bibinfo {author} {\bibfnamefont {In~Suk}\ \bibnamefont {Joung}}\ and\ \bibinfo {author} {\bibfnamefont {Thomas~E}\ \bibnamefont {Cheatham~III}},\ }\bibfield  {title} {\enquote {\bibinfo {title} {Determination of alkali and halide monovalent ion parameters for use in explicitly solvated biomolecular simulations},}\ }\href@noop {} {\bibfield  {journal} {\bibinfo  {journal} {The journal of physical chemistry B}\ }\textbf {\bibinfo {volume} {112}},\ \bibinfo {pages} {9020--9041} (\bibinfo {year} {2008})}\BibitemShut {NoStop}%
\bibitem [{\citenamefont {Burns}\ \emph {et~al.}(2017)\citenamefont {Burns}, \citenamefont {Faver}, \citenamefont {Zheng}, \citenamefont {Marshall}, \citenamefont {Smith}, \citenamefont {Vanommeslaeghe}, \citenamefont {MacKerell}, \citenamefont {Merz},\ and\ \citenamefont {Sherrill}}]{burns2017biofragment}%
  \BibitemOpen
  \bibfield  {author} {\bibinfo {author} {\bibfnamefont {Lori~A}\ \bibnamefont {Burns}}, \bibinfo {author} {\bibfnamefont {John~C}\ \bibnamefont {Faver}}, \bibinfo {author} {\bibfnamefont {Zheng}\ \bibnamefont {Zheng}}, \bibinfo {author} {\bibfnamefont {Michael~S}\ \bibnamefont {Marshall}}, \bibinfo {author} {\bibfnamefont {Daniel~GA}\ \bibnamefont {Smith}}, \bibinfo {author} {\bibfnamefont {Kenno}\ \bibnamefont {Vanommeslaeghe}}, \bibinfo {author} {\bibfnamefont {Alexander~D}\ \bibnamefont {MacKerell}}, \bibinfo {author} {\bibfnamefont {Kenneth~M}\ \bibnamefont {Merz}}, \ and\ \bibinfo {author} {\bibfnamefont {C~David}\ \bibnamefont {Sherrill}},\ }\bibfield  {title} {\enquote {\bibinfo {title} {The biofragment database (bfdb): An open-data platform for computational chemistry analysis of noncovalent interactions},}\ }\href@noop {} {\bibfield  {journal} {\bibinfo  {journal} {The Journal of chemical physics}\ }\textbf {\bibinfo {volume} {147}} (\bibinfo {year} {2017})}\BibitemShut {NoStop}%
\bibitem [{\citenamefont {Eastman}\ \emph {et~al.}(2023)\citenamefont {Eastman}, \citenamefont {Behara}, \citenamefont {Dotson}, \citenamefont {Galvelis}, \citenamefont {Herr}, \citenamefont {Horton}, \citenamefont {Mao}, \citenamefont {Chodera}, \citenamefont {Pritchard}, \citenamefont {Wang} \emph {et~al.}}]{eastman2023spice}%
  \BibitemOpen
  \bibfield  {author} {\bibinfo {author} {\bibfnamefont {Peter}\ \bibnamefont {Eastman}}, \bibinfo {author} {\bibfnamefont {Pavan~Kumar}\ \bibnamefont {Behara}}, \bibinfo {author} {\bibfnamefont {David~L}\ \bibnamefont {Dotson}}, \bibinfo {author} {\bibfnamefont {Raimondas}\ \bibnamefont {Galvelis}}, \bibinfo {author} {\bibfnamefont {John~E}\ \bibnamefont {Herr}}, \bibinfo {author} {\bibfnamefont {Josh~T}\ \bibnamefont {Horton}}, \bibinfo {author} {\bibfnamefont {Yuezhi}\ \bibnamefont {Mao}}, \bibinfo {author} {\bibfnamefont {John~D}\ \bibnamefont {Chodera}}, \bibinfo {author} {\bibfnamefont {Benjamin~P}\ \bibnamefont {Pritchard}}, \bibinfo {author} {\bibfnamefont {Yuanqing}\ \bibnamefont {Wang}},  \emph {et~al.},\ }\bibfield  {title} {\enquote {\bibinfo {title} {Spice, a dataset of drug-like molecules and peptides for training machine learning potentials},}\ }\href@noop {} {\bibfield  {journal} {\bibinfo  {journal} {Scientific Data}\ }\textbf {\bibinfo {volume} {10}},\ \bibinfo {pages} {11} (\bibinfo {year}
  {2023})}\BibitemShut {NoStop}%
\bibitem [{\citenamefont {Verstraelen}\ \emph {et~al.}(2016)\citenamefont {Verstraelen}, \citenamefont {Vandenbrande}, \citenamefont {Heidar-Zadeh}, \citenamefont {Vanduyfhuys}, \citenamefont {Van~Speybroeck}, \citenamefont {Waroquier},\ and\ \citenamefont {Ayers}}]{verstraelen2016minimal}%
  \BibitemOpen
  \bibfield  {author} {\bibinfo {author} {\bibfnamefont {Toon}\ \bibnamefont {Verstraelen}}, \bibinfo {author} {\bibfnamefont {Steven}\ \bibnamefont {Vandenbrande}}, \bibinfo {author} {\bibfnamefont {Farnaz}\ \bibnamefont {Heidar-Zadeh}}, \bibinfo {author} {\bibfnamefont {Louis}\ \bibnamefont {Vanduyfhuys}}, \bibinfo {author} {\bibfnamefont {Veronique}\ \bibnamefont {Van~Speybroeck}}, \bibinfo {author} {\bibfnamefont {Michel}\ \bibnamefont {Waroquier}}, \ and\ \bibinfo {author} {\bibfnamefont {Paul~W}\ \bibnamefont {Ayers}},\ }\bibfield  {title} {\enquote {\bibinfo {title} {Minimal basis iterative stockholder: atoms in molecules for force-field development},}\ }\href@noop {} {\bibfield  {journal} {\bibinfo  {journal} {Journal of Chemical Theory and Computation}\ }\textbf {\bibinfo {volume} {12}},\ \bibinfo {pages} {3894--3912} (\bibinfo {year} {2016})}\BibitemShut {NoStop}%
\bibitem [{\citenamefont {Rinaldi}\ \emph {et~al.}(2024)\citenamefont {Rinaldi}, \citenamefont {Bochkarev}, \citenamefont {Lysogorskiy},\ and\ \citenamefont {Drautz}}]{rinaldi2024charge}%
  \BibitemOpen
  \bibfield  {author} {\bibinfo {author} {\bibfnamefont {Matteo}\ \bibnamefont {Rinaldi}}, \bibinfo {author} {\bibfnamefont {Anton}\ \bibnamefont {Bochkarev}}, \bibinfo {author} {\bibfnamefont {Yury}\ \bibnamefont {Lysogorskiy}}, \ and\ \bibinfo {author} {\bibfnamefont {Ralf}\ \bibnamefont {Drautz}},\ }\bibfield  {title} {\enquote {\bibinfo {title} {Charge-constrained atomic cluster expansion},}\ }\href@noop {} {\bibfield  {journal} {\bibinfo  {journal} {arXiv preprint arXiv:2411.04062}\ } (\bibinfo {year} {2024})}\BibitemShut {NoStop}%
\bibitem [{\citenamefont {Dusson}\ \emph {et~al.}(2022)\citenamefont {Dusson}, \citenamefont {Bachmayr}, \citenamefont {Cs{\'a}nyi}, \citenamefont {Drautz}, \citenamefont {Etter}, \citenamefont {van~der Oord},\ and\ \citenamefont {Ortner}}]{dusson2022atomic}%
  \BibitemOpen
  \bibfield  {author} {\bibinfo {author} {\bibfnamefont {Genevieve}\ \bibnamefont {Dusson}}, \bibinfo {author} {\bibfnamefont {Markus}\ \bibnamefont {Bachmayr}}, \bibinfo {author} {\bibfnamefont {G{\'a}bor}\ \bibnamefont {Cs{\'a}nyi}}, \bibinfo {author} {\bibfnamefont {Ralf}\ \bibnamefont {Drautz}}, \bibinfo {author} {\bibfnamefont {Simon}\ \bibnamefont {Etter}}, \bibinfo {author} {\bibfnamefont {Cas}\ \bibnamefont {van~der Oord}}, \ and\ \bibinfo {author} {\bibfnamefont {Christoph}\ \bibnamefont {Ortner}},\ }\bibfield  {title} {\enquote {\bibinfo {title} {Atomic cluster expansion: Completeness, efficiency and stability},}\ }\href@noop {} {\bibfield  {journal} {\bibinfo  {journal} {Journal of Computational Physics}\ }\textbf {\bibinfo {volume} {454}},\ \bibinfo {pages} {110946} (\bibinfo {year} {2022})}\BibitemShut {NoStop}%
\bibitem [{\citenamefont {Hirshfeld}(1977)}]{hirshfeld1977bonded}%
  \BibitemOpen
  \bibfield  {author} {\bibinfo {author} {\bibfnamefont {Fred~L}\ \bibnamefont {Hirshfeld}},\ }\bibfield  {title} {\enquote {\bibinfo {title} {Bonded-atom fragments for describing molecular charge densities},}\ }\href@noop {} {\bibfield  {journal} {\bibinfo  {journal} {Theoretica chimica acta}\ }\textbf {\bibinfo {volume} {44}},\ \bibinfo {pages} {129--138} (\bibinfo {year} {1977})}\BibitemShut {NoStop}%
\bibitem [{\citenamefont {Zhu}\ and\ \citenamefont {Cheng}(2024)}]{zhu2024machine}%
  \BibitemOpen
  \bibfield  {author} {\bibinfo {author} {\bibfnamefont {Jia-Xin}\ \bibnamefont {Zhu}}\ and\ \bibinfo {author} {\bibfnamefont {Jun}\ \bibnamefont {Cheng}},\ }\bibfield  {title} {\enquote {\bibinfo {title} {Machine learning potential for electrochemical interfaces with hybrid representation of dielectric response},}\ }\href@noop {} {\bibfield  {journal} {\bibinfo  {journal} {arXiv preprint arXiv:2407.17740}\ } (\bibinfo {year} {2024})}\BibitemShut {NoStop}%
\bibitem [{\citenamefont {Zhang}\ \emph {et~al.}(2024)\citenamefont {Zhang}, \citenamefont {Andrade}, \citenamefont {Goldsmith}, \citenamefont {Raman}, \citenamefont {Li}, \citenamefont {Piaggi}, \citenamefont {Wu}, \citenamefont {Car},\ and\ \citenamefont {Selloni}}]{zhang2024electrical}%
  \BibitemOpen
  \bibfield  {author} {\bibinfo {author} {\bibfnamefont {Chunyi}\ \bibnamefont {Zhang}}, \bibinfo {author} {\bibfnamefont {Marcos~Calegari}\ \bibnamefont {Andrade}}, \bibinfo {author} {\bibfnamefont {Zachary~K}\ \bibnamefont {Goldsmith}}, \bibinfo {author} {\bibfnamefont {Abhinav~S}\ \bibnamefont {Raman}}, \bibinfo {author} {\bibfnamefont {Yifan}\ \bibnamefont {Li}}, \bibinfo {author} {\bibfnamefont {Pablo}\ \bibnamefont {Piaggi}}, \bibinfo {author} {\bibfnamefont {Xifan}\ \bibnamefont {Wu}}, \bibinfo {author} {\bibfnamefont {Roberto}\ \bibnamefont {Car}}, \ and\ \bibinfo {author} {\bibfnamefont {Annabella}\ \bibnamefont {Selloni}},\ }\bibfield  {title} {\enquote {\bibinfo {title} {Electrical double layer and capacitance of tio2 electrolyte interfaces from first principles simulations},}\ }\href@noop {} {\bibfield  {journal} {\bibinfo  {journal} {arXiv preprint arXiv:2404.00167}\ } (\bibinfo {year} {2024})}\BibitemShut {NoStop}%
\bibitem [{\citenamefont {Pozdnyakov}\ \emph {et~al.}(2020)\citenamefont {Pozdnyakov}, \citenamefont {Willatt}, \citenamefont {Bart{\'o}k}, \citenamefont {Ortner}, \citenamefont {Cs{\'a}nyi},\ and\ \citenamefont {Ceriotti}}]{pozdnyakov2020incompleteness}%
  \BibitemOpen
  \bibfield  {author} {\bibinfo {author} {\bibfnamefont {Sergey~N}\ \bibnamefont {Pozdnyakov}}, \bibinfo {author} {\bibfnamefont {Michael~J}\ \bibnamefont {Willatt}}, \bibinfo {author} {\bibfnamefont {Albert~P}\ \bibnamefont {Bart{\'o}k}}, \bibinfo {author} {\bibfnamefont {Christoph}\ \bibnamefont {Ortner}}, \bibinfo {author} {\bibfnamefont {G{\'a}bor}\ \bibnamefont {Cs{\'a}nyi}}, \ and\ \bibinfo {author} {\bibfnamefont {Michele}\ \bibnamefont {Ceriotti}},\ }\bibfield  {title} {\enquote {\bibinfo {title} {Incompleteness of atomic structure representations},}\ }\href@noop {} {\bibfield  {journal} {\bibinfo  {journal} {Physical Review Letters}\ }\textbf {\bibinfo {volume} {125}},\ \bibinfo {pages} {166001} (\bibinfo {year} {2020})}\BibitemShut {NoStop}%
\bibitem [{\citenamefont {Miura}\ \emph {et~al.}(2021)\citenamefont {Miura}, \citenamefont {Bartel}, \citenamefont {Goto}, \citenamefont {Mizuguchi}, \citenamefont {Moriyoshi}, \citenamefont {Kuroiwa}, \citenamefont {Wang}, \citenamefont {Yaguchi}, \citenamefont {Shirai}, \citenamefont {Nagao}, \citenamefont {Rosero‐Navarro}, \citenamefont {Tadanaga}, \citenamefont {Ceder},\ and\ \citenamefont {Sun}}]{Miura2021}%
  \BibitemOpen
  \bibfield  {author} {\bibinfo {author} {\bibfnamefont {Akira}\ \bibnamefont {Miura}}, \bibinfo {author} {\bibfnamefont {Christopher~J.}\ \bibnamefont {Bartel}}, \bibinfo {author} {\bibfnamefont {Yosuke}\ \bibnamefont {Goto}}, \bibinfo {author} {\bibfnamefont {Yoshikazu}\ \bibnamefont {Mizuguchi}}, \bibinfo {author} {\bibfnamefont {Chikako}\ \bibnamefont {Moriyoshi}}, \bibinfo {author} {\bibfnamefont {Yoshihiro}\ \bibnamefont {Kuroiwa}}, \bibinfo {author} {\bibfnamefont {Yongming}\ \bibnamefont {Wang}}, \bibinfo {author} {\bibfnamefont {Toshie}\ \bibnamefont {Yaguchi}}, \bibinfo {author} {\bibfnamefont {Manabu}\ \bibnamefont {Shirai}}, \bibinfo {author} {\bibfnamefont {Masanori}\ \bibnamefont {Nagao}}, \bibinfo {author} {\bibfnamefont {Nataly~Carolina}\ \bibnamefont {Rosero‐Navarro}}, \bibinfo {author} {\bibfnamefont {Kiyoharu}\ \bibnamefont {Tadanaga}}, \bibinfo {author} {\bibfnamefont {Gerbrand}\ \bibnamefont {Ceder}}, \ and\ \bibinfo {author} {\bibfnamefont {Wenhao}\ \bibnamefont {Sun}},\ }\bibfield
  {title} {\enquote {\bibinfo {title} {{Observing and Modeling the Sequential Pairwise Reactions that Drive Solid‐State Ceramic Synthesis}},}\ }\href@noop {} {\bibfield  {journal} {\bibinfo  {journal} {Advanced Materials}\ }\textbf {\bibinfo {volume} {33}} (\bibinfo {year} {2021})}\BibitemShut {NoStop}%
\bibitem [{\citenamefont {Gupta}\ \emph {et~al.}(2023)\citenamefont {Gupta}, \citenamefont {Yang},\ and\ \citenamefont {Ceder}}]{Gupta2023}%
  \BibitemOpen
  \bibfield  {author} {\bibinfo {author} {\bibfnamefont {Sunny}\ \bibnamefont {Gupta}}, \bibinfo {author} {\bibfnamefont {Xiaochen}\ \bibnamefont {Yang}}, \ and\ \bibinfo {author} {\bibfnamefont {Gerbrand}\ \bibnamefont {Ceder}},\ }\bibfield  {title} {\enquote {\bibinfo {title} {{What dictates soft clay-like lithium superionic conductor formation from rigid salts mixture}},}\ }\href@noop {} {\bibfield  {journal} {\bibinfo  {journal} {Nature Communications}\ }\textbf {\bibinfo {volume} {14}},\ \bibinfo {pages} {6884} (\bibinfo {year} {2023})}\BibitemShut {NoStop}%
\bibitem [{\citenamefont {Dai}\ \emph {et~al.}(2024)\citenamefont {Dai}, \citenamefont {Adhikari},\ and\ \citenamefont {Wen}}]{dai2024uncertainty}%
  \BibitemOpen
  \bibfield  {author} {\bibinfo {author} {\bibfnamefont {Jin}\ \bibnamefont {Dai}}, \bibinfo {author} {\bibfnamefont {Santosh}\ \bibnamefont {Adhikari}}, \ and\ \bibinfo {author} {\bibfnamefont {Mingjian}\ \bibnamefont {Wen}},\ }\bibfield  {title} {\enquote {\bibinfo {title} {Uncertainty quantification and propagation in atomistic machine learning},}\ }\href@noop {} {\bibfield  {journal} {\bibinfo  {journal} {arXiv preprint arXiv:2405.02461}\ } (\bibinfo {year} {2024})}\BibitemShut {NoStop}%
\bibitem [{\citenamefont {Zhu}\ \emph {et~al.}(2023)\citenamefont {Zhu}, \citenamefont {Batzner}, \citenamefont {Musaelian},\ and\ \citenamefont {Kozinsky}}]{zhu2023fast}%
  \BibitemOpen
  \bibfield  {author} {\bibinfo {author} {\bibfnamefont {Albert}\ \bibnamefont {Zhu}}, \bibinfo {author} {\bibfnamefont {Simon}\ \bibnamefont {Batzner}}, \bibinfo {author} {\bibfnamefont {Albert}\ \bibnamefont {Musaelian}}, \ and\ \bibinfo {author} {\bibfnamefont {Boris}\ \bibnamefont {Kozinsky}},\ }\bibfield  {title} {\enquote {\bibinfo {title} {Fast uncertainty estimates in deep learning interatomic potentials},}\ }\href@noop {} {\bibfield  {journal} {\bibinfo  {journal} {The Journal of Chemical Physics}\ }\textbf {\bibinfo {volume} {158}} (\bibinfo {year} {2023})}\BibitemShut {NoStop}%
\bibitem [{\citenamefont {Gal}\ and\ \citenamefont {Ghahramani}(2016)}]{gal2016_MC_dropout}%
  \BibitemOpen
  \bibfield  {author} {\bibinfo {author} {\bibfnamefont {Yarin}\ \bibnamefont {Gal}}\ and\ \bibinfo {author} {\bibfnamefont {Zoubin}\ \bibnamefont {Ghahramani}},\ }\bibfield  {title} {\enquote {\bibinfo {title} {A theoretically grounded application of dropout in recurrent neural networks},}\ }\href@noop {} {\bibfield  {journal} {\bibinfo  {journal} {Advances in neural information processing systems}\ }\textbf {\bibinfo {volume} {29}} (\bibinfo {year} {2016})}\BibitemShut {NoStop}%
\bibitem [{\citenamefont {Amini}\ \emph {et~al.}(2020)\citenamefont {Amini}, \citenamefont {Schwarting}, \citenamefont {Soleimany},\ and\ \citenamefont {Rus}}]{amini2020deep}%
  \BibitemOpen
  \bibfield  {author} {\bibinfo {author} {\bibfnamefont {Alexander}\ \bibnamefont {Amini}}, \bibinfo {author} {\bibfnamefont {Wilko}\ \bibnamefont {Schwarting}}, \bibinfo {author} {\bibfnamefont {Ava}\ \bibnamefont {Soleimany}}, \ and\ \bibinfo {author} {\bibfnamefont {Daniela}\ \bibnamefont {Rus}},\ }\bibfield  {title} {\enquote {\bibinfo {title} {Deep evidential regression},}\ }\href@noop {} {\bibfield  {journal} {\bibinfo  {journal} {Advances in neural information processing systems}\ }\textbf {\bibinfo {volume} {33}},\ \bibinfo {pages} {14927--14937} (\bibinfo {year} {2020})}\BibitemShut {NoStop}%
\bibitem [{\citenamefont {Wiberg}\ and\ \citenamefont {Rablen}(1993)}]{wiberg1993comparison}%
  \BibitemOpen
  \bibfield  {author} {\bibinfo {author} {\bibfnamefont {Kenneth~B}\ \bibnamefont {Wiberg}}\ and\ \bibinfo {author} {\bibfnamefont {Paul~R}\ \bibnamefont {Rablen}},\ }\bibfield  {title} {\enquote {\bibinfo {title} {Comparison of atomic charges derived via different procedures},}\ }\href@noop {} {\bibfield  {journal} {\bibinfo  {journal} {Journal of Computational Chemistry}\ }\textbf {\bibinfo {volume} {14}},\ \bibinfo {pages} {1504--1518} (\bibinfo {year} {1993})}\BibitemShut {NoStop}%
\bibitem [{\citenamefont {Marenich}\ \emph {et~al.}(2012)\citenamefont {Marenich}, \citenamefont {Jerome}, \citenamefont {Cramer},\ and\ \citenamefont {Truhlar}}]{marenich2012charge}%
  \BibitemOpen
  \bibfield  {author} {\bibinfo {author} {\bibfnamefont {Aleksandr~V}\ \bibnamefont {Marenich}}, \bibinfo {author} {\bibfnamefont {Steven~V}\ \bibnamefont {Jerome}}, \bibinfo {author} {\bibfnamefont {Christopher~J}\ \bibnamefont {Cramer}}, \ and\ \bibinfo {author} {\bibfnamefont {Donald~G}\ \bibnamefont {Truhlar}},\ }\bibfield  {title} {\enquote {\bibinfo {title} {Charge model 5: An extension of hirshfeld population analysis for the accurate description of molecular interactions in gaseous and condensed phases},}\ }\href@noop {} {\bibfield  {journal} {\bibinfo  {journal} {Journal of chemical theory and computation}\ }\textbf {\bibinfo {volume} {8}},\ \bibinfo {pages} {527--541} (\bibinfo {year} {2012})}\BibitemShut {NoStop}%
\bibitem [{\citenamefont {Mulliken}(1955)}]{mulliken1955electronic}%
  \BibitemOpen
  \bibfield  {author} {\bibinfo {author} {\bibfnamefont {Robert~S}\ \bibnamefont {Mulliken}},\ }\bibfield  {title} {\enquote {\bibinfo {title} {Electronic population analysis on lcao--mo molecular wave functions. i},}\ }\href@noop {} {\bibfield  {journal} {\bibinfo  {journal} {The Journal of chemical physics}\ }\textbf {\bibinfo {volume} {23}},\ \bibinfo {pages} {1833--1840} (\bibinfo {year} {1955})}\BibitemShut {NoStop}%
\bibitem [{\citenamefont {Gonze}\ and\ \citenamefont {Lee}(1997)}]{gonze1997dynamical}%
  \BibitemOpen
  \bibfield  {author} {\bibinfo {author} {\bibfnamefont {Xavier}\ \bibnamefont {Gonze}}\ and\ \bibinfo {author} {\bibfnamefont {Changyol}\ \bibnamefont {Lee}},\ }\bibfield  {title} {\enquote {\bibinfo {title} {Dynamical matrices, born effective charges, dielectric permittivity tensors, and interatomic force constants from density-functional perturbation theory},}\ }\href@noop {} {\bibfield  {journal} {\bibinfo  {journal} {Physical Review B}\ }\textbf {\bibinfo {volume} {55}},\ \bibinfo {pages} {10355} (\bibinfo {year} {1997})}\BibitemShut {NoStop}%
\bibitem [{\citenamefont {Wolverton}\ and\ \citenamefont {Zunger}(1998)}]{Wolverton1998_LiCoO2}%
  \BibitemOpen
  \bibfield  {author} {\bibinfo {author} {\bibfnamefont {C.}~\bibnamefont {Wolverton}}\ and\ \bibinfo {author} {\bibfnamefont {Alex}\ \bibnamefont {Zunger}},\ }\bibfield  {title} {\enquote {\bibinfo {title} {{First-principles prediction of vacancy order-disorder and intercalation battery voltages in Li$_x$CoO$_2$}},}\ }\href@noop {} {\bibfield  {journal} {\bibinfo  {journal} {Physical Review Letters}\ }\textbf {\bibinfo {volume} {81}},\ \bibinfo {pages} {606--609} (\bibinfo {year} {1998})}\BibitemShut {NoStop}%
\bibitem [{\citenamefont {Raebiger}\ \emph {et~al.}(2008)\citenamefont {Raebiger}, \citenamefont {Lany},\ and\ \citenamefont {Zunger}}]{raebiger2008charge}%
  \BibitemOpen
  \bibfield  {author} {\bibinfo {author} {\bibfnamefont {Hannes}\ \bibnamefont {Raebiger}}, \bibinfo {author} {\bibfnamefont {Stephan}\ \bibnamefont {Lany}}, \ and\ \bibinfo {author} {\bibfnamefont {Alex}\ \bibnamefont {Zunger}},\ }\bibfield  {title} {\enquote {\bibinfo {title} {Charge self-regulation upon changing the oxidation state of transition metals in insulators},}\ }\href@noop {} {\bibfield  {journal} {\bibinfo  {journal} {Nature}\ }\textbf {\bibinfo {volume} {453}},\ \bibinfo {pages} {763--766} (\bibinfo {year} {2008})}\BibitemShut {NoStop}%
\bibitem [{\citenamefont {Walsh}\ \emph {et~al.}(2018)\citenamefont {Walsh}, \citenamefont {Sokol}, \citenamefont {Buckeridge}, \citenamefont {Scanlon},\ and\ \citenamefont {Catlow}}]{Walsh2018}%
  \BibitemOpen
  \bibfield  {author} {\bibinfo {author} {\bibfnamefont {Aron}\ \bibnamefont {Walsh}}, \bibinfo {author} {\bibfnamefont {Alexey~A.}\ \bibnamefont {Sokol}}, \bibinfo {author} {\bibfnamefont {John}\ \bibnamefont {Buckeridge}}, \bibinfo {author} {\bibfnamefont {David~O.}\ \bibnamefont {Scanlon}}, \ and\ \bibinfo {author} {\bibfnamefont {C.~Richard~A.}\ \bibnamefont {Catlow}},\ }\bibfield  {title} {\enquote {\bibinfo {title} {{Oxidation states and ionicity}},}\ }\href@noop {} {\bibfield  {journal} {\bibinfo  {journal} {Nature Materials}\ }\textbf {\bibinfo {volume} {17}},\ \bibinfo {pages} {958--964} (\bibinfo {year} {2018})}\BibitemShut {NoStop}%
\bibitem [{\citenamefont {Vandermause}\ \emph {et~al.}(2022)\citenamefont {Vandermause}, \citenamefont {Xie}, \citenamefont {Lim}, \citenamefont {Owen},\ and\ \citenamefont {Kozinsky}}]{Vandermause2022_active}%
  \BibitemOpen
  \bibfield  {author} {\bibinfo {author} {\bibfnamefont {Jonathan}\ \bibnamefont {Vandermause}}, \bibinfo {author} {\bibfnamefont {Yu}~\bibnamefont {Xie}}, \bibinfo {author} {\bibfnamefont {Jin~Soo}\ \bibnamefont {Lim}}, \bibinfo {author} {\bibfnamefont {Cameron~J.}\ \bibnamefont {Owen}}, \ and\ \bibinfo {author} {\bibfnamefont {Boris}\ \bibnamefont {Kozinsky}},\ }\bibfield  {title} {\enquote {\bibinfo {title} {{Active learning of reactive Bayesian force fields applied to heterogeneous catalysis dynamics of H/Pt}},}\ }\href@noop {} {\bibfield  {journal} {\bibinfo  {journal} {Nature Communications}\ }\textbf {\bibinfo {volume} {13}},\ \bibinfo {pages} {5183} (\bibinfo {year} {2022})}\BibitemShut {NoStop}%
\bibitem [{\citenamefont {Ong}\ \emph {et~al.}(2013)\citenamefont {Ong}, \citenamefont {Richards}, \citenamefont {Jain}, \citenamefont {Hautier}, \citenamefont {Kocher}, \citenamefont {Cholia}, \citenamefont {Gunter}, \citenamefont {Chevrier}, \citenamefont {Persson},\ and\ \citenamefont {Ceder}}]{Ong2013_pmg}%
  \BibitemOpen
  \bibfield  {author} {\bibinfo {author} {\bibfnamefont {Shyue~Ping}\ \bibnamefont {Ong}}, \bibinfo {author} {\bibfnamefont {William~Davidson}\ \bibnamefont {Richards}}, \bibinfo {author} {\bibfnamefont {Anubhav}\ \bibnamefont {Jain}}, \bibinfo {author} {\bibfnamefont {Geoffroy}\ \bibnamefont {Hautier}}, \bibinfo {author} {\bibfnamefont {Michael}\ \bibnamefont {Kocher}}, \bibinfo {author} {\bibfnamefont {Shreyas}\ \bibnamefont {Cholia}}, \bibinfo {author} {\bibfnamefont {Dan}\ \bibnamefont {Gunter}}, \bibinfo {author} {\bibfnamefont {Vincent~L.}\ \bibnamefont {Chevrier}}, \bibinfo {author} {\bibfnamefont {Kristin~A.}\ \bibnamefont {Persson}}, \ and\ \bibinfo {author} {\bibfnamefont {Gerbrand}\ \bibnamefont {Ceder}},\ }\bibfield  {title} {\enquote {\bibinfo {title} {{Python Materials Genomics (pymatgen): A robust, open-source python library for materials analysis}},}\ }\href@noop {} {\bibfield  {journal} {\bibinfo  {journal} {Computational Materials Science}\ }\textbf {\bibinfo {volume} {68}},\ \bibinfo
  {pages} {314--319} (\bibinfo {year} {2013})}\BibitemShut {NoStop}%
\bibitem [{\citenamefont {Perdew}\ \emph {et~al.}(1996)\citenamefont {Perdew}, \citenamefont {Burke},\ and\ \citenamefont {Ernzerhof}}]{perdew1996_GGA}%
  \BibitemOpen
  \bibfield  {author} {\bibinfo {author} {\bibfnamefont {John~P}\ \bibnamefont {Perdew}}, \bibinfo {author} {\bibfnamefont {Kieron}\ \bibnamefont {Burke}}, \ and\ \bibinfo {author} {\bibfnamefont {Matthias}\ \bibnamefont {Ernzerhof}},\ }\bibfield  {title} {\enquote {\bibinfo {title} {Generalized gradient approximation made simple},}\ }\href@noop {} {\bibfield  {journal} {\bibinfo  {journal} {Physical review letters}\ }\textbf {\bibinfo {volume} {77}},\ \bibinfo {pages} {3865} (\bibinfo {year} {1996})}\BibitemShut {NoStop}%
\bibitem [{\citenamefont {Grimme}\ \emph {et~al.}(2010)\citenamefont {Grimme}, \citenamefont {Antony}, \citenamefont {Ehrlich},\ and\ \citenamefont {Krieg}}]{grimme2010_D3}%
  \BibitemOpen
  \bibfield  {author} {\bibinfo {author} {\bibfnamefont {Stefan}\ \bibnamefont {Grimme}}, \bibinfo {author} {\bibfnamefont {Jens}\ \bibnamefont {Antony}}, \bibinfo {author} {\bibfnamefont {Stephan}\ \bibnamefont {Ehrlich}}, \ and\ \bibinfo {author} {\bibfnamefont {Helge}\ \bibnamefont {Krieg}},\ }\bibfield  {title} {\enquote {\bibinfo {title} {A consistent and accurate ab initio parametrization of density functional dispersion correction (dft-d) for the 94 elements h-pu},}\ }\href@noop {} {\bibfield  {journal} {\bibinfo  {journal} {The Journal of chemical physics}\ }\textbf {\bibinfo {volume} {132}} (\bibinfo {year} {2010})}\BibitemShut {NoStop}%
\end{thebibliography}
\end{document}